%% file: main_icwsm.tex
\def\year{2022}\relax
\documentclass[letterpaper]{article} 
\usepackage{aaai25}  
\usepackage{times}  
\usepackage{helvet}  
\usepackage{courier}  
\usepackage[hyphens]{url}  
\usepackage{graphicx} 
\usepackage{natbib}  
\usepackage{caption} 
\DeclareCaptionStyle{ruled}{labelfont=normalfont,labelsep=colon,strut=off} 
\usepackage{algorithm}
\usepackage{algorithmic}
\usepackage{color,soul,xcolor,colortbl, graphicx,tabularx,enumitem,longtable,booktabs}
\usepackage{chngcntr,subcaption,varwidth,subfiles,csquotes}
\newcommand{\answerYes}[1]{\textcolor{blue}{#1}} 
\newcommand{\answerNo}[1]{\textcolor{teal}{#1}} 
\newcommand{\answerNA}[1]{\textcolor{gray}{#1}} 
 
\usepackage{newfloat}
\usepackage{listings}
\floatstyle{ruled}
\newfloat{listing}{tb}{lst}{}
\floatname{listing}{Listing}
\newcommand\question[1]{\textit{\enquote{#1}}}

\title{Large-Scale Analysis of Online Questions\\ Related to Opioid Use Disorder on Reddit}
\author{Tanmay Laud$^{1,2\footnote{Work done while student at UC San Diego}}$, Akadia Kacha-Ochana$^{3}$, Steven A. Sumner$^{3}$, Vikram Krishnasamy$^{3}$, Royal Law$^{4}$, Lyna Schieber$^{5}$, Munmun De Choudhury$^{6}$, and Mai ElSherief$^{7}$}
\affiliations{
\normalsize{$^{1}$Hippocratic AI,}\\
\normalsize{$^{2}$UC San Diego, San Diego, CA, USA},\\
\normalsize{$^{3}$Office of Strategy and Innovation, National Center for Injury Prevention and Control, }
\normalsize{Centers for Disease Control and Prevention, Atlanta, GA, USA,}\\

\normalsize{$^{4}$Division of Injury Prevention, National Center for Injury Prevention and Control, }
\normalsize{Centers for Disease Control and Prevention, Atlanta, GA, USA,}\\
\normalsize{$^{5}$Division of Overdose Prevention, National Center for Injury Prevention and Control, }
\normalsize{Centers for Disease Control and Prevention, Atlanta, GA, USA,}\\
\normalsize{$^{6}$Georgia Institute of Technology, Atlanta, GA, USA},\\
\normalsize{$^{7}$Northeastern University, Boston, MA, USA},\\

}

\usepackage{bibentry}

\begin{document}

\maketitle
\input{SectionFiles/abstract}
\input{SectionFiles/introduction}

\input{SectionFiles/related_work}

\input{SectionFiles/all_material}
\input{SectionFiles/results}

\input{SectionFiles/discussion}
\input{SectionFiles/conclusion}

{\footnotesize	
\bibliography{references}}
\input{SectionFiles/checklist}
\appendix
\input{SectionFiles/appendix.tex}

\end{document}

%% file: SectionFiles/abstract.tex
\begin{abstract}
Opioid use disorder (OUD) is a leading health problem that affects individual well-being as well as general public health. Due to a variety of reasons, including the stigma faced by people using opioids, online communities for recovery and support were formed on different social media platforms. In these communities, people share their experiences and solicit information by asking questions to learn about opioid use and recovery. However, these communities do not always contain clinically verified information. 
In this paper, we study natural language questions asked in the context of OUD-related discourse on Reddit. We adopt transformer-based question detection along with hierarchical clustering across 19 subreddits to identify six coarse-grained categories and 69 fine-grained categories of OUD-related questions.
Our analysis uncovers ten areas of information seeking from Reddit users in the context of OUD: drug sales, specific drug-related questions, OUD treatment, drug uses, side effects, withdrawal, lifestyle, drug testing, pain management and others, during the study period of 2018-2021. Our work provides a major step in improving the understanding of OUD-related questions people ask unobtrusively on Reddit. We finally discuss technological interventions and public health harm reduction techniques based on the topics of these questions.
\end{abstract}

%% file: SectionFiles/introduction.tex
\section{Introduction}
Opioid use disorder (OUD), a substance use disorder, is a problematic pattern of opioid use that causes significant impairment or distress~\cite{cdcopioid:2021}. OUD can lead to overdose or death. In 2021, opioid overdoses accounted for over 100,000 deaths in the United States alone~\cite{ahmad2021provisional}. 
In 2017, the U.S. Department of Health and Human Services declared the overdose crisis a public health emergency~\cite{hhs}. 
Despite the increased attention paid to the prevention of opioid overdose, death rates are at an all-time high. OUD is a complex public health problem that encompasses a variety of substances, both prescription and illicit.



People with OUD can be stigmatized on multiple levels. First, individuals with OUD can be perceived as dangerous, of moral failure, and labeled by pejorative terms such as ``addicts''~\cite{mclaren2023trends}. In response to the public stigma, people with OUD can experience internalized stigma and distress which is detrimental to their health. Finally, anticipated stigma is another layer of stigma experienced when a person with OUD is aware of these negative societal attitudes and develops expectations of being rejected. All the layers of stigma lead to people with OUD seeking information and social support through social media outlets that provide anonymity and a less judgmental environment~\cite{yao2020detection}. In particular, people with OUD use communities on Reddit to discuss substance misuse~\cite{Balsamo_Bajardi_De_Francisci_Morales_Monti_Schifanella_2023}, alternate treatments~\cite{chancellor2019discovering}, and recovery attempts~\cite{bunting2021socially}.
In these online spaces, individuals can freely share their experiences, ask questions, and receive support during recovery. 


Scholars have investigated content in online OUD communities, shedding light on various aspects of this pressing public health issue. Prior studies examined the prevalence of alternative treatments for OUD recovery~\cite{chancellor2016quantifying}, conducted a thematic analysis of posts about buprenorphine-naloxone~\cite{buprenorphineNLP}, defined event categories for characterizing information-seeking in OUD social discourse~\cite{sharif2023characterizing}, and developed computational methods to detect misinformation~\cite{misinfolargescale,elsherief2024identification} related to medications for OUD. A significant amount of content in online communities is structured in a Question and Answer (Q\&A) format~\cite{mamykina,asaduzzaman,hong2020,fu2018}. Despite these efforts, there remains a significant gap in understanding the types of questions individuals pose regarding OUD on social media platforms. This gap is crucial, as it speaks to the informational needs and uncertainties within the online OUD support communities, underscoring the necessity for further investigation into this domain. 
 This study bridges the gap by detecting, analyzing, and identifying the nature of questions posed by individuals with OUD through an analysis of 204,559 posts on Reddit. Understanding the nature of these questions will enable the design of public health interventions as well as opioid overdose prevention training guided by signals of pervasive questions and misconceptions on social media.

Given the ever-increasing volume of social media data, it is essential to develop computational techniques that can scale with the size of the data~\cite{socialmediachallenges}. Various hybrid approaches using Natural Language Processing (NLP) and machine learning have been proposed in the literature ~\cite{reddittextmining,opioidtextmining,hybridsystem} to study drug use on platforms like Reddit and Twitter. Recent studies have also attempted to computationally analyze the prevalence of OUD misinformation on popular platforms like Reddit, YouTube, Twitter and Drugs-Forum~\cite{misinfolargescale}. However, to the best of our knowledge, no known studies focus on the nature of questions asked by users engaging in OUD-related discussions.

\vspace{0.05in} \noindent\textbf{Our Contributions.} Public health communicators struggle to understand the extent of discussions around opioids by individuals who use substances and what questions are of greatest importance to such individuals ~\cite{volkow2016opioid,d2017social}. To address this concern, the present work performs the first large-scale analysis of opioid-related questions, drawing on advances in NLP and unsupervised learning.
We apply a mixed methods strategy to highlight the questions asked by people who use opioids followed by  topical analysis of these questions to identify the key areas that allow public health communicators to prioritize their efforts to address opioid related questions and potential misinformation. Our contributions are threefold:

\textbf{1.} \textit{We propose a NLP-based framework for the detection of questions in unstructured and noisy social media text.}

\textbf{2.} \textit{We implement this framework on 19 expert-curated OUD-related communities (subreddits) on Reddit to uncover the nature and topics of emerging questions on OUD.}

\textbf{3.} \textit{Based on our findings, we provide recommendations for public health policies and online moderation strategies.}

We partner with public health experts throughout the process. The experts, coauthors on this manuscript, are staff at the U.S. Centers for Disease Control and Prevention (CDC) with diverse backgrounds ranging from clinical medicine to expertise in health services research, data science, and program administration relevant to substance use. First, we develop an unsupervised learning approach to automatically detect questions from 204,559 posts. Next, we build a meaningful summary of each post that explains the post in less than 100 words. We combine the summaries and questions and perform dense semantic clustering. Finally, we present a qualitative analysis by highlighting the key questions being asked by people who used opioids. We discuss design and public health implications of this work. 

\vspace{0.05in} \noindent\textbf{Ethics and Privacy.} The data used for this research was publicly available when downloaded, and there was no interaction between the authors and social media users. Consequently, this work did not qualify for approval from the relevant Institutional Review Board. To protect the privacy and anonymity of the users, we conducted all the analysis on firewalled servers and did not upload code/data on any public cloud. 
In addition, we made minor edits to the reported user questions to avoid re-identification and reduce the traceability of users. We did not process any Reddit user-specific or demographic fields from Reddit in our analysis so that a user's race, sex, age, or other private attributes are protected.

%% file: SectionFiles/related_work.tex
\section{Related Work}


\subsection{Studying OUD Support Forums}

Social media provides online forums for people to make connections that may influence attitudes and behaviors. Various analyses have been conducted on these forums to uncover these behaviors ~\citep{maclean2015forum77, kim2017scaling, rubya2017video, d2017social} and monitor unethical activities ~\citep{sarker2016social, sarker2020mining}. For example, \citet{drugAbuseTwitter} analyzed Twitter data to observe the conversation and engagement of these networks with regard to prescription drug misuse. They found that Twitter users who discussed prescription drug misuse online are surrounded by others who also discuss it, potentially reinforcing the negative behavior. Young people and pregnant women have been found to be the most vulnerable to misinformation related to the dissemination of opioid use practices~\cite{browniehigh,liang2021identifying}. Further, social media platforms have been identified as a potential source of illegal promotion of the sale of controlled substances directly to consumers~\cite{drugAbuseLinkTwit, sequeira2019large}. In the case of cannabinoids, for instance, research has identified content that describes, encourages, promotes~\cite{cannabinoids}, and normalizes the consumption of illicit substances~\cite{tweetAnalysis, epidiopioids}. The topics deduced by our pipeline underscores these findings at a larger scale on recent data, highlighting the current gaps in online toxicovigilance.

\subsection{Analyzing Reddit in View of OUD}

Reddit has been a preferred social media platform in the study of OUD experiences, disclosures, and support seeking or provisioning~\citep{lu2019investigate}. \citet{routePatterns} developed a word embedding based procedure to find alternative terms referring to opioid, their routes of administration, and drug-tampering methods. 
\citet{alternateTreatments} provided the first large-scale Reddit based study of alternative treatments for OUD recovery. They adopted machine learning across 63 subreddits to precisely identify posts related to opioid recovery and discovered potential alternative treatments. 
\citet{buprenorphineNLP} conducted a thematic analysis of posts about buprenorphine-naloxone from the subreddit \textit{r/suboxone}. They applied NLP to generate meta-information and curate samples of salient posts. 
\citet{stigmaSources} examined the sources of stigma that people seek support for on Reddit by comparing posts with and without stigma keywords and identified the type (condition, intervention) and source (provider-based, public, self, structural) of stigma.
\cite{sharif2023characterizing} was the first to define event categories for characterizing information-seeking on Reddit where they focused on \textit{r/suboxone} subreddit. Notably, they observed low quality annotation through crowdsourcing the categories and eventually relied on in-house experts. We draw inspiration from these past works and apply a larger scale deep learning aided process with experts-in-the-loop to uncover topic trends specifically surrounding OUD questions.



\subsection{Analysis of Q\&A on Social Media}
The high volume of data generated via Q\&A websites is an essential asset to understanding how the members share knowledge, how they behave and how useful these websites are for them. While there have been studies on many Q\&A websites, the majority of the effort has been towards Stack Overflow\footnote{\url{https://insights.stackoverflow.com/survey/2020}}. \citet{mamykina} showed that Stack Overflow was more effective at getting questions answered than other Q\&A websites with high response rates and fast response times. \citet{asaduzzaman} explored why questions go unanswered and \citet{fastAns} identified factors that lead to fast answers. Researchers have examined user opinions \citep{linares}, pain-points \citep{cummaudo}, topics \citep{rosen,barua,beyer2020}, and overall behavior \citep{sadowksi}.  Studies have also investigated methods of improving Stack Overflow, such as helping users find the information \citep{nadi2020,zhang2019stackoverflow}, identifying expert users and using their knowledge to help the community \citep{ford2018}. 

Fewer studies explored the content of other Q\&A communities. \citet{hong2020} have analyzed how users share knowledge on a health Q\&A website, and \cite{fu2018} analyzed the quality of answers on social Q\&A websites. Other studies have provided ways of improving Q\&A websites, such as increasing the overall politeness \cite{wang2021}, increasing the popularity of academic answers \citep{Zhang2019ScientificKC}, finding experts to answer questions \citep{Shen2020HelpingTI,procaci2019}, and improving the way the community welcomes new users \citep{santos2020}. 

The effectiveness of hierarchical clustering in capturing complex structures in textual data is supported by \citet{hu-etal-2019-diachronic}, who proposed a framework using deep contextualized embeddings to model fine-grained word senses and their temporal evolution. Additionally, \citet{Ding_2024} introduced a multi-step reasoning framework leveraging prompt-based large language models to analyze social media language patterns and their association with national health outcomes. Their study demonstrates the potential of advanced language models in extracting nuanced patterns from social media discussions, which can inform public health strategies. Furthermore, \citet{Romano_Sharif_Basak_Gatto_Preum_2024} developed a theme-driven keyphrase extraction framework tailored for social media, designed to capture clinically relevant keyphrases from user-generated health texts. This approach facilitates the identification of salient concepts within specific themes, enhancing the analysis of social media discourse. 

These studies collectively highlight the importance of advanced modeling techniques in capturing the hierarchical and dynamic nature of textual information, thereby enhancing the effectiveness of clustering and topic modeling approaches.

%% file: SectionFiles/all_material.tex
\section{Data}

We leveraged Reddit as a platform for our study due to the array of advantages it embodies. Reddit is a growing social media and content curation
 site with over 300 million monthly active users and over 100,000 active communities.~\footnote{\url{https://foundationinc.co/lab/reddit-statistics/}}
Additionally, Reddit does not have a restrictive character limit (10,000 characters for a comment and 40,000 for a post) for the posts so users can provide detailed information. The posts (called ``submissions" in Reddit) are organized in communities that discuss specific substances, such as (\textit{r/Methadone}, \textit{r/naltrexone}, \textit{r/oxycodone}, \textit{r/kratom}) allowing for ease of identification on forums related to opioids. For example, the subreddit r/suboxone describes itself as \textit{`A subreddit created to provide a place for discussion on Suboxone and other forms of buprenorphine, welcome to all whether it be short-term or long-term usage for MAT, for pain management'} and has 20.4k members. Finally, Reddit provides an important source of social support and mutual aid for persons who use opioids. Prior work indicates that online social support networks are beneficial to persons who use opioids, particularly during events where isolation from other social support resources may occur~\cite{opioidcovid}.


We used the Pushshift API to collect Reddit posts from January 2018 to September 2021~\cite{baumgartner2020pushshift}. Prior work identified 21 subreddits where OUD discussions take place~\cite{misinfolargescale}. Public health co-authors of this paper inspected and validated the subreddits' relevance to OUD. We identified 19 subreddits relevant to our study: \textit{r/Carfentanil}, \textit{r/fentanyl}, \textit{r/heroin}, \textit{r/heroinaddiction}, \textit{r/HeroinHeroines}, \textit{r/kratom}, \textit{r/lean}, \textit{r/loperamide}, \textit{r/Methadone}, \textit{r/naltrexone}, \textit{r/oxycodone}, \textit{r/opiates}, \textit{r/OpiatesRecovery}, \textit{r/OpiateChurch}, \textit{r/OurOverUsedVeins}, \textit{r/Opiatewithdrawal}, \textit{r/quittingkratom}, \textit{r/suboxone}, \textit{r/Tianeptine}. We dropped r/modquittingkratom and r/opiatesmemorial from our analysis since these subreddits contain very few posts (4-10 samples) and are inactive.

We filtered the data for posts with text only and excluded posts marked as `[removed]'/`[deleted]' by the API. We collected a total of 204,559 submissions at the end of the filtering process. Table~\ref{table:1} summarizes the descriptive statistics of the dataset. The number of posts per subreddit ranged from 43 to 90,015, with a mean of 14,336 posts, median of 3787, and a standard deviation of 23,306 posts. We present macro-statistics such as mean text length in characters, mean number of tokens, and mean sentence count in Appendix Table~\ref{table:macro}.


\begin{table}[!ht]
\centering
\resizebox{\columnwidth}{!}{%
\begin{tabular}{llll}
\hline
\rowcolor[HTML]{C0C0C0} 
Unique subreddits                          & 19                           &                                                        &        \\ \hline
\multicolumn{1}{l|}{Total Posts}           & \multicolumn{1}{l|}{408,398} & \multicolumn{1}{l|}{Avg Text Length} & 509.87 \\ \hline
\multicolumn{1}{l|}{Non-Text posts} & \multicolumn{1}{l|}{135,981} & \multicolumn{1}{l|}{Avg No. of Tokens} & 97.27 \\ \hline
\multicolumn{1}{l|}{Deleted/removed posts} & \multicolumn{1}{l|}{67,858}  & \multicolumn{1}{l|}{Avg. No. of Sents.}      & 6.4    \\ \hline
\multicolumn{1}{l|}{Text posts}            & \multicolumn{1}{l|}{204,559} & 
\\ \hline
\end{tabular}
}
\caption{General statistics of the Reddit dataset obtained via Pushshift API for 19 subreddits relevant to OUD. The period of study is January 2018 to September 2021.}
\label{table:1}
\end{table}

\section{Methods}


\subsection{Data Processing Pipeline}
\begin{figure*}[h!]
    \centering
    \includegraphics[scale=0.29]{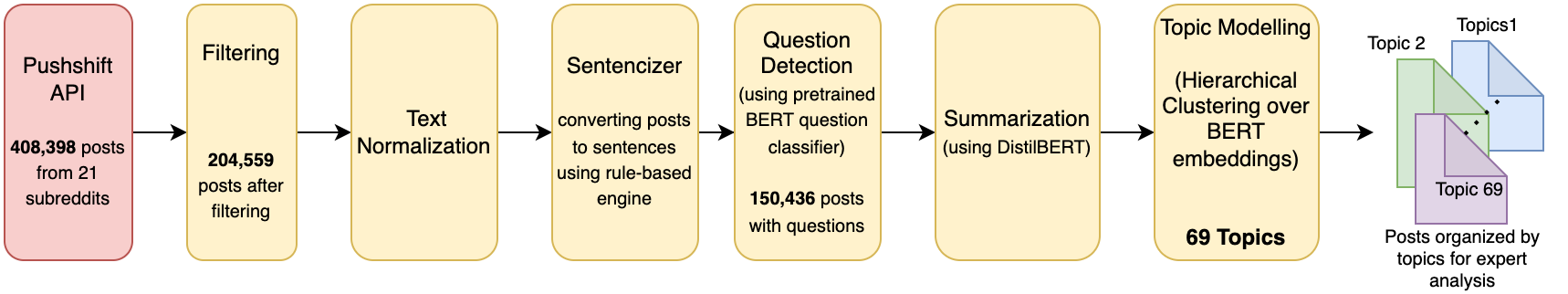}
    \caption{Data Processing Pipeline. We first preprocess Reddit posts by normalizing text and correcting errors, then using a deep neural network to detect and filter questions. Then, we cluster the model generated summary of these posts into meaningful topics using hierarchical clustering.}
    \label{fig:data_proc}
\end{figure*}
Due to a lack of labeled text data, we constructed an unsupervised learning based  processing pipeline. Firstly, we pre-processed our Reddit data by filtering to focus on text-only posts. Then, text-only posts were normalized. Text normalization allows us to expand lexical contractions and correct common spelling errors. We
processed the data for correcting errors like typos, expanding contractions and abbreviations, and unconventional spellings.
 We then detect the questions using a deep neural network classifier and filter out posts that do not contain any questions. The questions and a machine generated extractive summary of the text are used to then cluster the posts into meaningful topics using hierarchical clustering. The natural language processing components are explained in detail as follows. 

\noindent
\textbf{Text Normalization.}\label{section:normalization} Social media language is comprised of errors like typos, abbreviations, unconventional spellings, phonetic substitutions, and other lexical variants. One approach to normalizing social media text is to preprocess posts to produce a more standard rendering of these lexical variants. For example, \textit{`se u 2morw!!!'} could be normalized to \textit{`see you tomorrow!'}. Such a normalization approach is useful as a preprocessing step for applications that rely on keyword match or word frequency statistics, such as topic analysis.
We tokenized the data using NLTK~\cite{nltk}.

To preprocess the text, we used the Lexnorm package~\cite{lexnorm} to remove URLs, email addresses, and personal pronouns. Lexnorm is specifically designed for medical social media data which closely relates to the domain of our subreddits. 
Contractions with apostrophes were expanded (eg. `don't' $\rightarrow$ `do not') and subsequently, those without apostrophes were expanded based on a dictionary. Domain-specific abbreviations were expanded with a lexicon of domain-specific abbreviations expansions using Lexnorm. For spelling corrections, the package provides an unsupervised spelling correction module that uses the corpus to construct a set of probable correction candidates based on relative token frequency 
and edit distance threshold. 
After normalizing the text, we perform sentence segmentation using a rule-based spacy model\footnote{\url{https://spacy.io/api/sentencizer}. The following is an example of a normalized text from the dataset.
}

\textbf{Raw Text:} \textit{\textbf{Ive} recently taken interest in researching kratom and as a \textbf{noob} I wanted to run \textbf{smthing} by you guys}.

\textbf{Normalized Text:} \textit{\textbf{i have} recently taken interest in researching kratom and as a \textbf{someone who is new} i wanted to run \textbf{something} by you guys .}


\noindent\textbf{Question Detection.} The task of determining whether a sentence is a question is complicated in the case of Reddit posts since social media posts have irregular sentence formation and not every question is ended with a question mark. We use a finetuned BERT-based question classifier with a BERT-mini~\cite{bert-mini} backbone to detect whether each sentence in a post is a question.\footnote{The model weights were pre-initialized from the huggingface model~\cite{huggingface} {Question Detection Model Weights \\\url{https://huggingface.co/shahrukhx01/question-vs-statement-classifier}}. The model was trained on a Quora-based classification dataset~\footnote{\url{https://www.kaggle.com/datasets/stefanondisponibile/quora-question-keyword-pairs}} with a reported 99.7\% accuracy on Kaggle https://www.kaggle.com/code/shahrukhkhan/question-vs-statement-classification-mini-bert.} 
On average, the number of question marks per post is 0.78 (Appendix A1 Table~\ref{table:macro}). Compared to this naive approach of counting `?', the classifier detects 1.6 questions per post on average, indicating that we capture more nuanced questions using a trained model than a simple regex capture of `?'.
To assess the performance of the question detector on our domain, a random sample of 100 posts was annotated through crowdsourcing. Four annotators were tasked to label the posts indicating whether a question indicates a question or not. This process resulted in a Fleiss $\kappa$~\cite{fleiss1971measuring} of 0.77 indicating strong agreement (annotation details in Appendix A1). The naive question mark-based detector scored an accuracy of 68\% whereas our model scored an 80\% accuracy (12\% points improvement).

Additionally, we analyzed the position of a question in a post to determine if questions tend to appear at the very beginning of a post or the end. Appendix A1   Figure~\ref{fig:ques_pos} indicates that for most of the posts, the questions tend to appear at the end (59.4\% of posts harboring a question in the last three sentences).
We dropped posts from our analysis that did not contain any questions, resulting in a total of 150,436 posts for analysis.
The position of the question helps in identifying relatively which part of the post would need to be considered to establish a minimal baseline summary (explained in next section) of posts longer than 100 words on average.

\noindent
\textbf{Summarization.}\label{section:summary} After processing and normalizing the posts, we produced a short (max 100 tokens) summary of the posts for public health experts to utilize for any downstream task-specific labeling. This was achieved using a distilled version RoBERTa trained on summarization task~\cite{bertsum}. We also prepared a baseline summary of each post to contrast with the model summary. The baseline is constructed by extracting the question in the post and a maximum of two sentences before the question. If there were multiple questions in a post, we repeat the procedure for every question and concatenate the outputs. We calculated the BLEU~\cite{bleu} and ROUGE~\cite{rouge} scores of the model summary relative to our baseline and report the results in Appendix A1 Table~\ref{table:summary}. Our ROUGE scores range between 0.41-0.86 indicating moderate to strong overlap.

\noindent\textbf{Topic Modeling.} Next, we looked at categorizing the posts by topics, or clustering. To cluster the posts into interpretable topics, we followed a multi-step process. 
First, in order to get a meaningful vector representation of the text, we concatenated the title, model-generated summary, and the extracted questions to form a single passage per post and generate its embedding through a sentence embedding model. The title and post text provide additional context especially when the questions are incomplete and reference topics in the posts. For example, for a question like `how would it help?', the `it' can be inferred only on the basis of the post content. Prior work leveraged titles as the target summaries for text summarization models to learn from~\cite{huggingface_text_summ}. Thus, the titles act as a high level summary while the summary of the post captures salient fine-grained information from the post. We employed a Mini-LM model~\cite{reimers-2019-sentence-bert}, a compute efficient and high performance model, for generating 384 dimensional sentence embeddings for each post. Since we only use the title and summary (maximum of 100 tokens), we are able to process longer text posts without any length challenges.

\noindent
\noindent\textbf{Compression.} The dimension of the sentence embedding is 384 which is considerably high for running a clustering algorithm in reasonable time given the size of our dataset and the curse of dimensionality. Hence, we compressed the data to a lower dimensional space such that the embeddings still preserved a similar locality structure as that of the higher dimensional space. We used UMAP~\cite{umap} algorithm which is a general purpose dimension reduction technique. We set the number of neighbours to 15 and metric as cosine to get 5-dimensional embedding vectors.

\noindent
\textbf{Clustering.} Real-world textual information is not grouped under single independent domains but is rather composed in a hierarchy of domains. Classic topic modeling is unable to model the correlations among these hierarchies because of a single distribution over topics in each document. Therefore, hierarchical clustering~\cite{hierclust} has been used recently for topic modelling on social media text~\cite{elsherief2024identification}. One of the popular algorithms for this task is HDBScan~\cite{hdbscan}.
HDBScan requires a minimum cluster size as a hyperparameter. In order to find the best value for this parameter, we ran a search over multiple values between 3 and 600 \footnote{We ran the search over the values (3, 5, 7, 9, 12, 15, 20, 25, 30, 50, 100, 150, 200, 220, 240, 250, 260, 500, 600) with euclidean metric} with the goal of optimizing the average coherence score (UMass Coherence)\footnote{The UMass Coherence score was calculated using gensim library~\cite{gensim}}. We found the optimum cluster size to be 200 corresponding to 69 clusters (Refer to Appendix A1 Figure~\ref{fig:umass_min_clus} for details). 
Additionally, we conducted a rapid qualitative analysis~\cite{vindrola2020rapid} of the clustering results of the embedding. The five public health expert reviewers concluded that the models performed well and messages assigned to each topic were appropriate, indicating strong coherence of the approach.

\noindent\textbf{Topic Interpretation.} In order to find the top relevant words in the cluster to explain the topic of that cluster, we computed the TF-IDF~\cite{tfidf} score of each word with respect to the documents in its cluster. Then, we used the Maximal Marginal Relevance score to get the top 4 keywords per topic, using the BERTopic~\cite{bertopic} python library. Appendix A2 Table~\ref{table:topicinterpret} gives examples of a few generated topics from the corpus.

%% file: SectionFiles/results.tex
\section{Results}
\begin{figure*}
    \centering
\begin{subfigure}{\columnwidth}
    \includegraphics[scale=0.20]{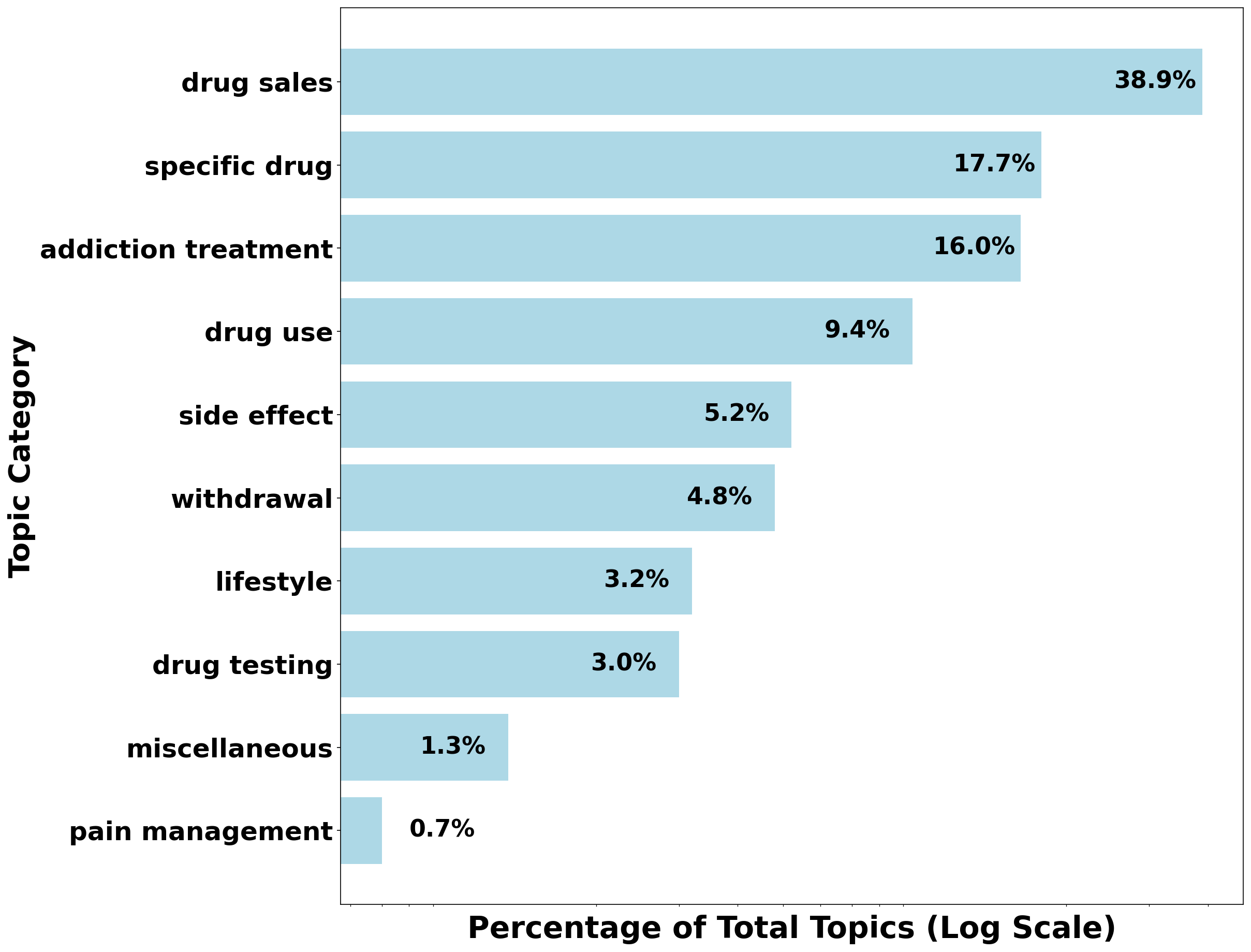}
    \caption{Distribution of Topic Groups}
    \label{fig:topicgroup}
\end{subfigure}
\begin{subfigure}{\columnwidth}
    \centering
    \includegraphics[scale=0.20]{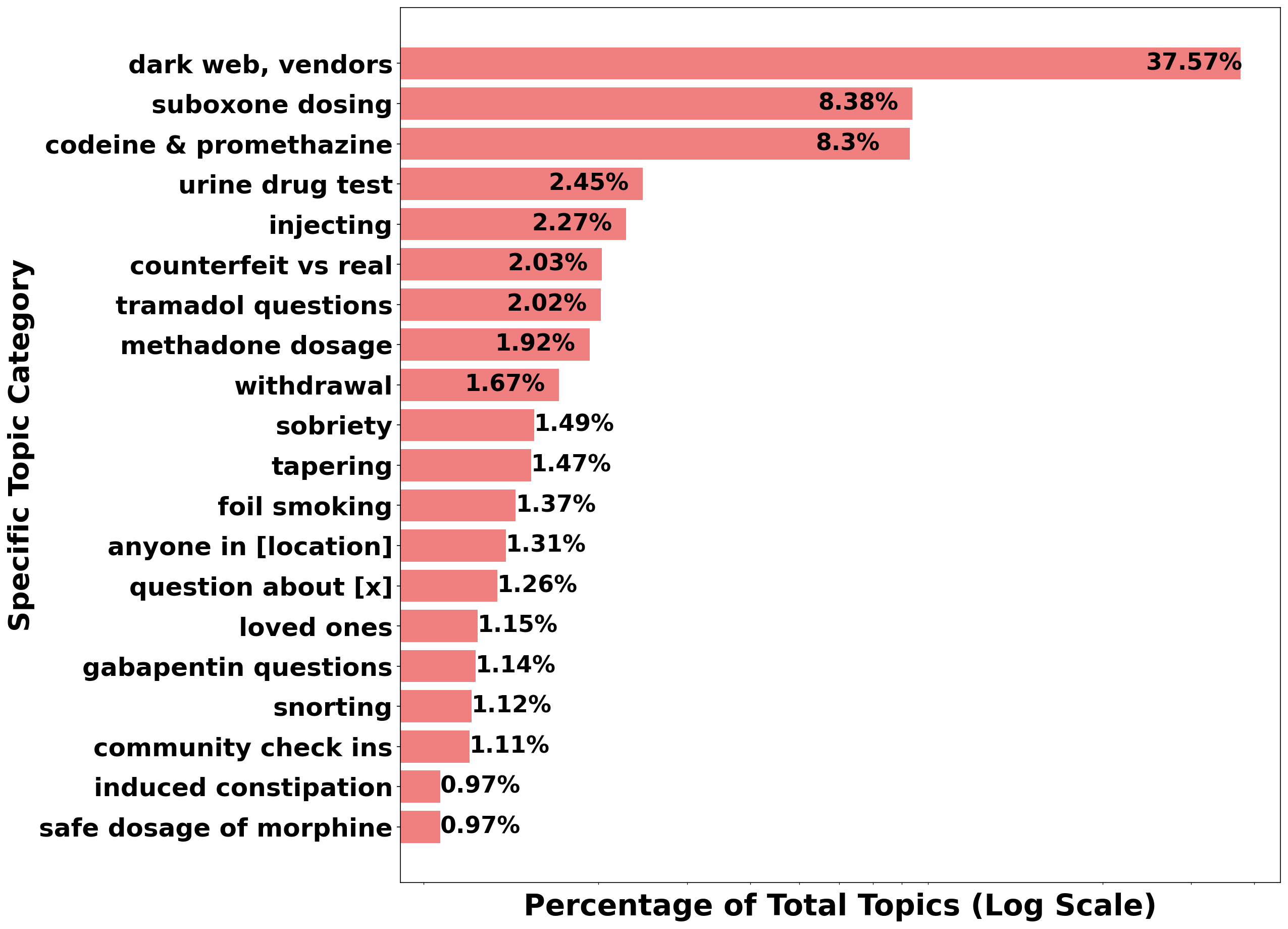}
    \caption{Distribution of Top 20 topics}
    \label{fig:top20topics}
\end{subfigure}
\caption{Topics categorized into groups by public health experts based on the top 100 relevant posts in every topic. Drug sales accounts for the highest volume by topic groups (a) which is also evident in the top 20 topics (b) where topics of the dark web and vendors have the highest counts.}
\end{figure*}
Our experiments with the clustering models led to the identification of 69 topic clusters. For each of these topics, we extracted 100 posts that were closest to their respective cluster centroids.
Our public health authors gave each topic a well defined name guided by the most salient keywords given by the topic modelling pipeline and the questions asked in the posts. 
To assess the performance of the model-based clustering process, two reviewers coded 207 randomly sampled posts (3 posts per each of the 69 topics). Each reviewer independently assessed the proportion of posts correctly assigned to a given topic, yielding similar results (95.2\% and 96.1\%). The inter-rater agreement between the reviewers as calculated via Cohen's Kappa was 0.72 indicating substantial agreement. Appendix A2 Table~\ref{tab:topicgroups} shows the 69 individual topics of interest. For ease of display and discussion, the 69 topics were then grouped by the public health coauthors into 10 higher level categories
: drug sales, specific drug, addiction treatment, drug use, side effect, withdrawal, lifestyle, drug testing, pain management, and miscellaneous. The majority of the topics fall under the `drug sales' group (38.9\%) followed by topics on `specific drugs' (17.7\%) and `addiction treatment' (16\%). Figure~\ref{fig:topicgroup} shows the overall distribution of the 10 categories. Further, when we look at the most popular individual topics (Figure~\ref{fig:top20topics}), we see that most of the social media chatter is about the purchase and sale of substances (37.6\%). This topic surpasses the number of posts that ask questions about suboxone dosing, the second most popular topic and an approved treatment for OUD (8.4\%), by a large margin. 

\subsection{Major Topics}
Example posts are shared and briefly discussed below for the 7 most prevalent topics out of the 69 topics identified via unsupervised learning.

\noindent\textbf{Questions about Dark Web.} A large proportion of users asked questions about how they can procure opioids, such as through illicit online sales. One user asked, \question{I just read the rules about how dark web how tos. but as the rules said , there are specific reddits for that. So could someone point me in a trustworthy reddit page for that?}. Another user asked about Empire, a dark web market: \question{For those that browse the onions? is empire a solid site?}
The impact of COVID-19 on the Dark Web has also been noted, with anonymization techniques enabling illegal operations and posing risks to individuals seeking to bypass IP monitoring mechanisms~\cite{coviddarkweb}.
For instance, one user wanted to know how to setup a Virtual Private Network (VPN) service:
\question{Can someone walk me through how to use to VPN on iPhone and can anyone explain how to use the onion?}
The COVID-19 pandemic led to lockdowns in many locations making it difficult for individuals who use substances to access in-person networks for purchasing substances and increasing reliance on the Dark Net Market (DNM):
\question{Buying on the DNM. Safe during these ones of Corona?	Was thinking of buying off the dnm. anyone else had any issues during these times?} The pandemic has affected drug utilization trends, raising concerns about the safety of purchasing opioids during these times~\cite{covidtrends}.

\noindent\textbf{Questions about Suboxone dosing.}
The questions in this topic mostly pertained to timing and dosing of suboxone based treatment for OUD, whether alone or in combination with other substances. The timing and dosing of suboxone-based treatment for opioid use disorder (OUD) are critical considerations for individuals seeking effective management~\cite{buprenorphineTapering}. 
Evidently, we have questions about timing:
\question{if i took 4mg of suboxone at 8a.m. this morning when will it be cool to do opiates again?
it is been 24 hours since i last used any fent. should i be okay to take a sub or should i wait longer.?}, and about quantity: \question{do you know how many days do you have to taper from 21mg of suboxone to 0? i want to know about the fastest tapering?}
In the context of combining suboxone with opioid pain relievers, it is essential to consider potential interactions and adverse effects~\cite{kratomMixing}. 
Users inquired about mixing suboxone with opioid pain relievers like Percoret:
\question{Anyone been on low dose suboxone and ran out? We’re u okay? What would happen if you took a Percocet ten hours after ur last dose
took 12mg yesterday morning , is it worth it if i were to buy some dope? or should i wait longer
} Users also asked questions about combining naltrexone (another treatment of OUD) with suboxone:
\question{when can i take naltrexone after suboxone?	how soon can you take naltrexone after suboxone? i only took one sub over the last 5 days and this morning i ran out of suboxone and like to take naltrexone.}

\noindent\textbf{Questions about Promethazine and Codeine.}
Promethazine is an antihistamine primarily used to treat allergies and motion sickness. 
Codeine is an opioid pain reliever most commonly used to treat pain and to reduce coughing. It is available in combination with Acetaminophen or Aspirin and in some cough and cold medications. Combining the use of promethazine and opioids has been reported, including among individuals injecting drugs, as it is thought to potentiate the “high” from opioids~\cite{promethazine}. The growing popularity is evident from questions like:
\question{i have been searching for information on promethazine since i got a shitload of it for free. can it be used for anything positive or should i just get rid of it? hope any of you have experiences with it.} and 
\question{i have a ton of promethazine from chemotherapy and i read that if you combine it with opiates you get pretty high. any body know how much to mix? any information would be great thanks !}. 

Drug tampering practices, with the aim to enhance drug effects, are accessible on the Internet and are practiced by a growing number of people who use drugs~\cite{drugTampering2018}. One such process is Cold Water Extraction (CWE)~\cite{coldWater2020}, a technique used to extract a substance from a mixture via cold water; CWE came up frequently in the context of codeine:
\question{i usually have about 8 pills so is it safe to drink on 240mg of codeine? is it even safe to drink on any amount of codeine? if so how many beers/shots maximum.}
\question{How does separating codeine from acetaminophen work using cold water extraction?}

\noindent\textbf{Questions about urine drug testing.} Individuals who have to undergo drug testing either due to law enforcement action or job requirements tended to ask about how to achieve a negative result. As observed in previous studies, this includes pausing medication, excessive hydration with drinking water or detoxification beverages, or use of synthetic urine~\cite{urineTest}:
\question{I'm a daily user of a gram to maybe 1.5 a day and have a drug test on Friday (urine) how long will I have to stop for it to be out of my urine? I know it lasts longer for heavy users so I really need some answers here on what you guys think.}
\question{hey so i have been off oxys for about 11 days , just got hired at a new job. well i have to take a urine test in the next 24 hours. i am a skinny girl , will i pass? anyone had this happen and pass? any advice will be appreciated !}

\noindent\textbf{Questions about injecting.}
Prolonged drug use can lead to vein damage and sclerosis~\cite{ciccarone2017}, making it challenging to access veins for injection and it is crucial to consider the complications associated with this practice~\cite{Pieper2009}. 
Individuals who inject drugs frequently reported problems accessing veins and sought out information on injection advice. For instance,
\question{How do I find new veins? I smoked for about 3 years and switched to shooting. My first 2 weeks i kept missing and was just stabbing my veins and not injecting. Now I can't get any blood to draw back when I hit them.} Some ask about injection through veins in the legs instead of arms and the size of needles:
\question{i have been searching online for any kind of guide for finding the veins that can be injected into in the legs - beyond charts warning about all the arteries in the upper legs and describing groin injection. should i be using a different sized needle?}. 

\noindent\textbf{Questions about counterfeit drugs.} 
Counterfeit medicines are medicines that are deliberately and fraudulently mislabelled with respect to identity and/or source. According to the United States Drug Enforcement Administration, drug criminal networks are responsible for mass-producing fake prescription opioid pills. These counterfeit pills can be mixed with a potentially lethal dose of other drugs, such as fentanyl~\cite{donnell2023}. Some users asked about aspects of pills they purchased, to infer if they were fake or real, including color, shape, fragility, and if certain pills are still produced. One user asks the community to send pictures to corroborate the medication they purchased. \question{Does anyone get the Sandoz 40mg wafers? The ones with the E 132. If so can you send me couple pics front back side. Do they fake these? Thanks}. Another user inquired about aspects of the opioid pain reliever, Percocet. \question{my friend came back from mexico with some pink \textbf{perc 10s} but there was not anything on the pill and it was big like a \textbf{rp20} but it broke easily and the powder was white. does anybody know if these are real or not?} We note that users tend to use imprint text on pills and numbers indicating dosage to colloquially refer to certain medications in this type of questions. Examples include the following questions \question{have you all seen fake \textbf{rps}, even the \textbf{30s}?} and \question{are \textbf{alg 264s} soft and scratched up?} (both referencing Oxycodone Hydrochloride). This provides insights on question phrasing and how persons who use opioids refer to certain medications. 







\noindent\textbf{Questions about Tramadol.} Another emerging topic in the natural language of questions was related to the synthetic opioid Tramadol. Research on Tramadol has shown that those prescribed Tramadol have a higher risk of death in comparison to those prescribed anti-inflammatory medications~\cite{tramadolA, tramadolB}. The misuse of Tramadol has contributed to the overdose crisis~\cite{tramadolCrisis}. For this topic, users asked about building tolerance and higher dosages. \question{i am wondering why tramadol has given me a better high , and a better feeling when i take it , and also a better feeling the day after i have taken them. i have had hydros and its not the same. they are not as potent , and it comes on more harsh. is that because my tolerance is higher from tram? i have taken 200 - 250mg of tramadol before , felt great. and i have taken 25mg of hydro and it felt like a less strong and shorter lasting version of tramadol. any explanations or comments would be appreciated. : )} and \question{I'v built a tolerance to 500mg tramadol so	how do i get the magic again? do i just increase the dosage to 600 of 700+mg thanks}. Another user asked about the utility of Tramadol for the management of oxycodone withdrawal. \question{Can tramadol help u get off oxy}. 

\subsection{Major Topic Groups}
Table~\ref{tab:topicgroups} shows the distribution of the 69 topics when grouped via public health experts into larger categories or themes. Ten categories were ultimately identified, with drug sales-related questions being most prevalent, followed by questions about specific substances and addiction treatment. These categories provide a more macro-level perspecive on the main themes or questions being asked in the forums. Procuring substances, as noted previously, is a dominant theme and consistent with the large and worsening substance use and overdose crisis in the United States. Of note, addiction treatment related questions comprised the third most popular category. This is an encouraging finding as questions about treatment reflect demand for recovery services.

%% file: SectionFiles/discussion.tex
\section{Discussion}
This study sought to address a fundamental need in public health communication efforts--better understanding the questions that individuals who use opioids are asking on social media and what information they are seeking. To date, health and public health related communications largely have originated from expert-derived perspective and content. While valuable, many health experts may not have lived experience pertaining to substance use, which may adversely affect the relevance and utility of health information that is produced and disseminated. Indeed, owing to the difficulty in understanding and synthesizing large volumes of unstructured text data, it has  not been possible to quantitatively guide health experts based on large scale information from the perspectives of individuals who use substances. Our work fills this important gap by developing and applying computational methods in combination with public health expert review to assess large volumes of actual questions asked by individuals who use substances to more fully elucidate the questions and information that the platform users may be most interested in. An exploration of these themes has relevance for public health practice and prevention strategies.

Concerningly, we found that the largest share of questions discussed the procurement of substances, such as via dark web markets and similar modalities. The prevalent nature of these questions is a reflection of the ongoing opioid overdose crisis in the U.S. and the high burden of OUD among individuals. The majority of questions about drug procurement discussed topics such as the use of the dark web, virtual private networks including the ``onion" or ``tor" networks, and help to identify vendors. While many health professionals recognize that individuals may purchase substances online, the scale of these activities may not be fully appreciated.
This suggests that an area of increased relevance for communication is sharing appropriate cautions about risks associated with procuring substances. While some studies have begun to explore this area~\cite{li2021demystifying} there is a need to better understand how health professionals can best share prevention information with individuals who procure substances online. For example, recent research examined where users were procuring sodium nitrite, a substance used in suicide attempts, to help inform a discussion of possible prevention strategies~\cite{das2024emerging}.

Drugs, of a wide variety of classes, can be intentionally or unintentionally contaminated with highly lethal synthetic opioids, such as fentanyl. Individuals seeking substances procured from sources online highlights concerns about the complexity of overdose prevention in drug markets where the composition of substances cannot be readily ascertained. For this reason, harm reduction strategies such as the provision of fentanyl test strips, are being pursued by some communities as a means by which individuals can learn whether substances they intend to consume may contain lethal contaminants~\cite{peiper2019fentanyl}.

Over 17\% of questions posted were related to specific drugs, ranging from comparatively benign substances such as poppy seeds to illicit opioids such as heroin. Notably, many questions around specific drugs were related to emerging drugs such as kratom and tianeptine~\cite{marraffa2018poison}. Emerging substances are particularly prone to misinformation due to the lack of established information regarding the dangers of the substance from trusted communicators to address misinformation claims. Monitoring specific drug questions can inform health communicators to provide more timely and targeted guidance for specific and emerging drugs that may not yet be in the purview of public health. 

Only 16\% of posted questions referenced addiction treatment and consistent with the fact that utilization of MOUD continues to lag. In 2020, 11.2\% of people with opioid use disorder received MOUD~\cite{samhsa2020}. Further, access to care is not consistent across the US. Twenty-eight million Americans live more than ten miles from the nearest buprenorphine provider~\cite{langabeer2020geographic}. In addition, studies demonstrate that persons who use drugs seek out buprenorphine for treatment of withdrawal symptoms or self-treatment of OUD~\cite{gandhi2022alternative}. Evidence supports better OUD treatment outcomes among those who use non-prescribed buprenorphine prior to entering treatment~\cite{williams2022non}. Taken together, these issues may account for the posted questions referencing addiction treatment.

Questions around drug use made up the fourth most prevalent topic we identified, indicating that people who use substances do use online forums to seek out advice on using substances. There is growing attention in public health to harm reduction strategies for OUD, such as overdose education, naloxone distribution, syringe-services, and use of drug-checking supplies~\cite{sue2021bringing}, which can be promoted in online environments as a way to attempt to reduce the risk of death among individuals who are actively using substances.

\noindent\textbf{Implications for Design and Public Health.} A salient topic for this work is the pragmatic implications for public health and prevention efforts. While awareness of misinformation as a threat to health is growing, there remain practical knowledge gaps to guide the work of public health professionals. At a macro-level, helping to surface a quantitative awareness of what are the key questions that individuals who use substances are asking, helps to elucidate the areas where health misinformation may appear. This awareness is a first step in aiding public health professionals to combat misinformation. For example, misinformation on a given topic may become prominent if there is absence of official health information from trusted sources. This can be a particular challenge for new health topics in which scientific information is emerging, however, such knowledge can help to focus research and knowledge-generation efforts on these high priority areas. Additionally, while health officials may have a priori hypotheses about what are the primary areas for concern for health misinformation, quantitative approaches such as those presented in this manuscript can provide improved validation or adjustment of hypotheses, something that has not been heretofore possible.

Furthermore, methods such as those employed in this research can help public health and healthcare professionals better understand salient topics of interest to patients. For example, as shown in Table 5, the most prevalent topic within the drug use group was injection drug use. Injection drug use is of considerable significance given the elevated risk of acquiring or transmitting infectious diseases, including the human immunodeficiency virus (HIV), which can result in notable community outbreaks ~\cite{peters2016hiv}. Additionally, injection drug use can pose other unique health-related harms such as bloodstream infection, sepsis, and infective endocarditis~\cite{rudasill2019clinical}. Ongoing discussions about injection drug use online indicate continued need to work to prevent such use and provides an opportunity for enhanced education to occur in online environments. As an additional example of public health relevance, the side effect category was comprised of over 10 unique concerns arising from the use of substances. Pharmacovigilance, which is the science of detection of adverse events from medicines, is continually striving for earlier detection of such adverse events. While such work has historically focused on analyis of traditional clinical data, there is growing recognition of the value of online data to inform pharmacovigilance efforts and this work helps contribute to that growing body of evidence~\cite{sarker2015utilizing}.

This study underscored the significance of social media as a place for the public to discuss and share information related to opioids. Users seek out and place trust in information provided through online communities and sharing accurate health information on Reddit is a generally unexplored area for public health communicators. While some public health campaigns have explored the use of peer- or influencer-shared health information, much additional research is needed in understanding the most effective way to share health information related to substances with online communities~\cite{bonnevie2020using}. Indeed, public health programs are beginning to explore modalities as diverse as chatbots to peer influences in such efforts and further research is needed to fully understand the most trusted and effective approaches~\cite{bonnevie2021layla}. 

While Reddit already possesses strong forum moderation performed by volunteer moderators, there is a growing body of research working to understand how health professionals and others can support, empower, and educate such volunteer moderators~\cite{weld2022makes}. Prior research has shown that effective moderation on Reddit enhances the efficacy and safety of support provided online and extending and expanding such work to substance use is a rich area for further development~\cite{wadden2021effect}. Opportunities may exist for public health professionals to better support and engage with moderators over key topics to minimize misinformation. While mechanisms to support and facilitate this type of collaboration have not yet been concretely defined and studies are needed to assess the utility of novel models of user support, opportunities for bi-directional knowledge sharing could exist. For example, in addition to helping to educate moderators on health topics, health professionals can also benefit from knowledge of emerging challenges and complex clinical phenomenon that are discussed online and moderators may have early awareness~\cite{spadaro2022reddit}. Prior work has also shown that during the Covid-19 pandemic, social media companies have taken a coordinated approach to content moderation by directing users to reliable information from government healthcare agencies~\cite{baker2020covid19}. A similar approach could be implemented in the context of OUD subreddits if such specific reliable information addressing the topics that were discovered in this study was designed.

Our findings help social media designers in understanding the characteristics of each of the subreddits. Users can be provided recommendations that can help them in choosing on which community to post their questions. By directing questions to communities that can handle them more effectively, we can contribute to supporting and bolstering community well-being.

\noindent\textbf{Limitations and Future Work.} While this work performed the largest study to date on assessing questions posed about opioids online, there are some limitations to note. First, our analysis was limited to text data and did not explore other forms of expression such as communication through emojis or content shared in images. Additional work to incorporate signs from information beyond text-based data may help in identifying questions or topics discussed using coded or highly colloquial language or expressions. Additionally, this study focused only on Reddit data, although the framework we employed could be extended to other platforms such as Twitter. There are also emerging social audio communication channels for information sharing and longer-form group discussion, such as Discord and Clubhouse, that merit future attention as a potential space for sharing of health information. We did not explore a wide range of embedding sizes in our clustering analysis due to compute constraints and consider this as future work.
Furthermore, this study focused only on English language communications; extending this approach to multilingual data would help to benefit a broader group of individuals. As with any online research, new online communities or forums are continually being established and the robustness and ongoing comprehensiveness of this approach depends on identification of new and relevant subreddits. 
Moreover, within the field of overdose prevention, there are persistent questions (common across time) and more emergent questions. For example, there are a finite number of treatments for opioid use disorder and questions about the primary treatments are likely to remain relevant for longer periods of time~\cite{bell2020medication}. However, there also exists more temporally relevant topics, such as new substances that may be mixed with opioids and present novel questions related to health effects and overdose prevention~\cite{love2023opioid}. Distinguishing both types of questions is vital and we leave discriminating the persistent versus emerging topics research for future work. Lastly, this study focuses on the questions asked by users, however, future work would ideally explore the answers provided to these questions as an additional important step in fully understanding and correcting health misinfomation that exists.

%% file: SectionFiles/conclusion.tex
\section{Conclusion}
Our research develops and deploys a natural language processing and machine learning based pipeline combined with public health expert qualitative assessment to provide a large-scale assessment of questions asked about opioids on Reddit. This work is important as medical and public health officials have historically relied on expert opinion to guide what health information is most pressing and relevant to communicate to individuals who use substances. However, to improve an evidence-informed and data-driven strategy for public health communications, information is needed on what questions individuals who use substances are actually asking and robustly discussing. Our work demonstrates a viable framework for developing such information, which can ultimately be used to inform and improve the relevancy of public health communications focused on reducing opioid overdose. 

%% file: SectionFiles/checklist.tex
\section{Paper Checklist}

\begin{enumerate}

\item For most authors...
\begin{enumerate}
    \item  Would answering this research question advance science without violating social contracts, such as violating privacy norms, perpetuating unfair profiling, exacerbating the socio-economic divide, or implying disrespect to societies or cultures?
    \answerYes{Yes}
  \item Do your main claims in the abstract and introduction accurately reflect the paper's contributions and scope?
    \answerYes{Yes. Please refer to Methods, Results and Discussion section.}
   \item Do you clarify how the proposed methodological approach is appropriate for the claims made? 
    \answerYes{Yes, and we also discuss limitations of the approach in the Discussion section (Limitations and Future Work).}
   \item Do you clarify what are possible artifacts in the data used, given population-specific distributions?
    \answerYes{Yes. Data section}
  \item Did you describe the limitations of your work?
    \answerYes{Yes. Discussion section (Limitations and Future Work).}
  \item Did you discuss any potential negative societal impacts of your work?
    \answerYes{Yes.}
      \item Did you discuss any potential misuse of your work?
    \answerNo{No.}
    \item Did you describe steps taken to prevent or mitigate potential negative outcomes of the research, such as data and model documentation, data anonymization, responsible release, access control, and the reproducibility of findings?
    \answerYes{Yes.}
  \item Have you read the ethics review guidelines and ensured that your paper conforms to them?
    \answerYes{Yes}
\end{enumerate}

\item Additionally, if you ran machine learning experiments...
\begin{enumerate}
  \item Did you include the code, data, and instructions needed to reproduce the main experimental results (either in the supplemental material or as a URL)?
    \answerNo{No. Reddit recently changed their terms of service and does not allow distribution of their data as per our interpretation\footnote{\url{https://www.reddit.com/r/modnews/comments/134tjpe/reddit_data_api_update_changes_to_pushshift_access/}}. Future work will have to comply with the new Reddit DATA API.}
  \item Did you specify all the training details (e.g., data splits, hyperparameters, how they were chosen)?
    \answerNo{We did not train a new model but used various trained models for question detection and clustering of data. Please refer to Methods and Appendix.}
     \item Did you report error bars (e.g., with respect to the random seed after running experiments multiple times)?
    \answerNA{NA}
	\item Did you include the total amount of compute and the type of resources used (e.g., type of GPUs, internal cluster, or cloud provider)?
    \answerNo{No.}
     \item Do you justify how the proposed evaluation is sufficient and appropriate to the claims made? 
    \answerYes{Yes. Please refer to Results and Discussion sections.}
     \item Do you discuss what is ``the cost`` of misclassification and fault (in)tolerance?
    \answerNA{NA}
  
\end{enumerate}

\item Additionally, if you are using existing assets (e.g., code, data, models) or curating/releasing new assets, \textbf{without compromising anonymity}...
\begin{enumerate}
  \item If your work uses existing assets, did you cite the creators?
    \answerYes{Yes. We cite the relevant Reddit, Pushshift and Huggingface assets in various sections.}
  \item Did you mention the license of the assets?
    \answerNo{No.}
  \item Did you include any new assets in the supplemental material or as a URL?
    \answerNo{No.}
  \item Did you discuss whether and how consent was obtained from people whose data you're using/curating?
    \answerNA{NA}
  \item Did you discuss whether the data you are using/curating contains personally identifiable information or offensive content?
    \answerYes{Yes.}
\item If you are curating or releasing new datasets, did you discuss how you intend to make your datasets FAIR ?
\answerNA{NA}
\item If you are curating or releasing new datasets, did you create a Datasheet for the Dataset? 
\answerNA{NA}
\end{enumerate}

\end{enumerate}

%% file: SectionFiles/appendix.tex
 \pagebreak
\section*{Appendix}
\subsection{A1. Methods Appendix}\label{methods_app}


\subsubsection{Question Detection Annotation Guidelines}\label{annotation}
The annotation task was to label 100 randomly sampled posts whether they contained a question or not. The annotators were warned that the data might contain sensitive content. The selected crowdworkers from UpWork Platform have a data annotation background with an average of four years of experience and fluency in English language. 


\noindent\textbf{Task Guidelines:}
The task is to label each post as yes or no depending on the following:
1) \textit{Yes}: if any of the sentences in the post is a valid question
2) \textit{No}: if none of the sentences in the post is a valid question.

We indicated to the UpWork crowdworkers that questions may not necessarily have a question mark or have WH words like what, why, when, etc. Rhetorical statements like 'guess what?' are not be marked as questions even if they have a '?' in the text.

We designed the following qualification test that enabled us to assess the crowdworkers before working on our task. We designed these questions to mimic unstructured social media text that are similar to posts in our dataset. We only selected crowdworkers who scored 100\% on the qualification test and provided sound reasoning for each their responses, indicating a strong understanding of the task.
 
\noindent\textbf{Qualification Test:} The candidates were asked to label three posts as follows.

Does this text have a question (one which expects an answer from the receiver)?:

1)~\textit{Hey y'all! So, I was scrolling through my feed, and guess what I stumbled upon? The craziest cat video eva, this kitty was doing backflips or somethin! Couldn't stop laughin!} \textbf{Ground Truth: No. There is no question in the text. 'guess what I stumbled upon? is a rhetorical statement often used to draw attention.}

2)~\textit{yo fam, wassup? So, I was chillin' with my crew last night, right? And we were into this game drop. it's hella costly, should I cop it or nah? Like, is it worth droppin' mad cash on it?} \textbf{Ground Truth: Yes. The question is in the latter half of the text. The 'wassup?' is just an informal greeting.}

3)~\textit{Yo squad, peep this - last night's vibes were off the charts! We're all hyped about this new game drop, ya know? But the real talk, that price tag got me questioning if it's worth it.} \textbf{Ground Truth: No.}

After completing the qualification test, the selected annotators proceeded to annotate the 100 random samples provided. Next, we computed the ground truth labels based on majority voting strategy. We then contrasted the annotators' labels to a baseline that detects the presence of a ``?'' and the BERT-based \textit{question vs. statement} classifier\footnote{The model weights were pre-initialized from the huggingface model~\cite{huggingface} {Question Detection Model Weights \\\url{https://huggingface.co/shahrukhx01/question-vs-statement-classifier}}. The model was trained on a Quora-based classification dataset~\footnote{\url{https://www.kaggle.com/datasets/stefanondisponibile/quora-question-keyword-pairs}} with a reported 99.7\% accuracy on Kaggle https://www.kaggle.com/code/shahrukhkhan/question-vs-statement-classification-mini-bert.}. We find that the \textit{question vs. statement} classifier achieves the highest accuracy of 80\%, indicating its suitability for our domain.

\begin{table}[!h]
\resizebox{\columnwidth}{!}{%
\begin{tabular}{l|l|l|l|l|l|l}
\hline
\rowcolor[HTML]{9B9B9B}
\textbf{Subreddit} & \textbf{Text} & \textbf{No. of} & \textbf{No. of } & \textbf{No. of} & \textbf{Question \% } & \textbf{No. of} \\ 
\rowcolor[HTML]{9B9B9B}
 & \textbf{Length} & \textbf{Words} & \textbf{? Marks} & \textbf{Sentences} &  & \textbf{Questions} \\ \hline
Methadone & 794 & 153 & 1.27 & 8.97 & 63.04 & 1.82 \\ \hline
\rowcolor[HTML]{EFEFEF} 
naltrexone & 505 & 94 & 1.21 & 6.70 & 77.05 & 1.52 \\ \hline
Opiatewithdrawal & 813 & 158 & 1.16 & 10.45 & 62.93 & 1.75 \\ \hline
\rowcolor[HTML]{EFEFEF} 
suboxone & 674 & 130 & 1.15 & 8.16 & 64.77 & 1.70 \\ \hline
OurOverUsedVeins & 621 & 112 & 0.96 & 7.05 & 52.38 & 1.69 \\ \hline
\rowcolor[HTML]{EFEFEF} 
loperamide & 883 & 169 & 0.96 & 10.00 & 59.53 & 1.55 \\ \hline
OpiatesRecovery & 884 & 169 & 0.94 & 10.82 & 49.70 & 1.77 \\ \hline
\rowcolor[HTML]{EFEFEF} 
opiates & 460 & 88 & 0.90 & 5.73 & 58.52 & 1.57 \\ \hline
fentanyl & 382 & 75 & 0.86 & 4.85 & 59.07 & 1.53 \\ \hline
\rowcolor[HTML]{EFEFEF} 
heroin & 332 & 65 & 0.79 & 4.32 & 55.88 & 1.51 \\ \hline
kratom & 361 & 67 & 0.70 & 4.64 & 56.26 & 1.45 \\ \hline
\rowcolor[HTML]{EFEFEF} 
HeroinHeroines & 354 & 68 & 0.69 & 4.84 & 44.31 & 1.59 \\ \hline
Tianeptine & 255 & 46 & 0.68 & 3.46 & 58.85 & 1.44 \\ \hline
\rowcolor[HTML]{EFEFEF} 
OpiateChurch & 243 & 47 & 0.68 & 3.42 & 56.90 & 1.40 \\ \hline
quittingkratom & 713 & 136 & 0.61 & 9.30 & 39.82 & 1.58 \\ \hline
\rowcolor[HTML]{EFEFEF} 
lean & 143 & 26 & 0.54 & 2.10 & 56.67 & 1.26 \\ \hline
heroinaddiction & 377 & 73 & 0.38 & 4.46 & 33.85 & 1.45 \\ \hline
\rowcolor[HTML]{EFEFEF} 
Carfentanil & 550 & 91 & 0.26 & 5.79 & 30.23 & 1.15 \\ \hline
oxycodone & 32 & 5 & 0.06 & 1.41 & 19.09 & 1.10 \\ \hline
\rowcolor[HTML]{EFEFEF} 
\end{tabular}%
}
\caption{\textbf{Macro Statistics.} The reported statistics are the average across the posts for each subreddit. The text length refers to the number of characters in a post. Question \% is the proportion of questions in sentences.}
\label{table:macro}
\end{table}

\begin{figure}[!h]
    \centering
    \includegraphics[width = 8cm, height = 5cm]{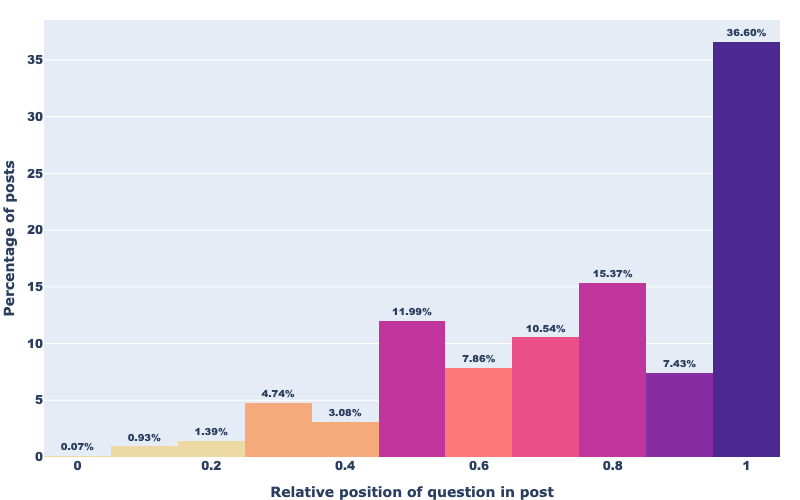}
    \caption{\textbf{Relative position of question in a post on average.} Most of the posts have questions asked at the end of post.}
    \label{fig:ques_pos}
\end{figure}

\begin{table}
\centering
\begin{tabular}{l|r|r}
\hline
\rowcolor[HTML]{9B9B9B}
\textbf{Subreddit} & \multicolumn{1}{l|}{\textbf{BLEU Score}} & \multicolumn{1}{l}{\textbf{ROUGE Score}} \\ \hline
r/Methadone        & 0.2911  & 0.4478 \\ \hline\rowcolor[HTML]{EFEFEF} 
r/naltrexone       & 0.4135  & 0.5289 \\ \hline
r/Opiatewithdrawal & 0.3001  & 0.4312 \\ \hline\rowcolor[HTML]{EFEFEF} 
r/suboxone         & 0.3396  & 0.4867 \\ \hline
r/OurOverUsedVeins & 0.4268  & 0.6002 \\ \hline\rowcolor[HTML]{EFEFEF} 
r/loperamide       & 0.3710  & 0.4939 \\ \hline
r/OpiatesRecovery  & 0.2628  & 0.4102 \\ \hline\rowcolor[HTML]{EFEFEF} 
r/opiates          & 0.4006  & 0.5655 \\ \hline
r/fentanyl         & 0.4356  & 0.5888 \\ \hline\rowcolor[HTML]{EFEFEF} 
r/heroin           & 0.4586  & 0.6252 \\ \hline
r/kratom           & 0.4296  & 0.6134 \\ \hline\rowcolor[HTML]{EFEFEF} 
r/HeroinHeroines   & 0.3913  & 0.6025 \\ \hline
r/Tianeptine       & 0.4643  & 0.6678 \\ \hline\rowcolor[HTML]{EFEFEF} 
r/OpiateChurch     & 0.4962  & 0.6498 \\ \hline
r/quittingkratom   & 0.2938  & 0.4609 \\ \hline\rowcolor[HTML]{EFEFEF} 
r/lean             & 0.5662  & 0.7464 \\ \hline
r/heroinaddiction  & 0.2612  & 0.6218 \\ \hline\rowcolor[HTML]{EFEFEF}
r/Carfentanil      & 0.3563  & 0.5459 \\ \hline 
r/oxycodone        & 0.5821  & 0.8599 \\ \hline\rowcolor[HTML]{EFEFEF} 

\end{tabular}
\caption{\textbf{Summarization Performance of Model relative to Baseline.} The baseline summary is constructed by extracting the question in the post and a maximum of two sentences before the post. If there were multiple questions in a post, then the result is concatenated for each question. We calculated the BLEU~\cite{bleu} and ROUGE~\cite{rouge} scores of the model generated summary relative to our baseline.
}
\label{table:summary}
\end{table}

\begin{figure*}[!h]
\centering
\begin{subfigure}{\columnwidth}
  \centering
  \includegraphics[scale=0.5]{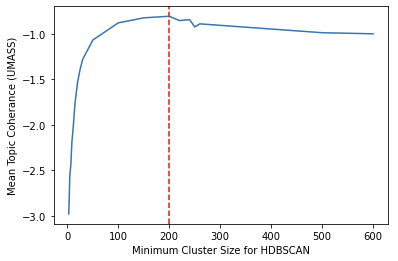}
  \caption{Mean Topic Coherence vs Min Cluster Size}
  \label{fig:umass_min_clus}
\end{subfigure}%
\begin{subfigure}{\columnwidth}
  \centering
  \includegraphics[scale=0.5]{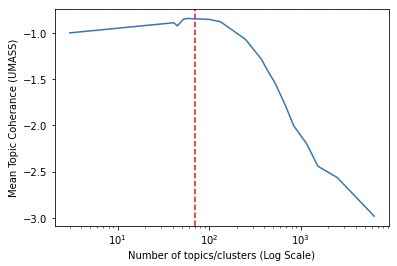}
  \caption{Mean Topic Coherence vs Number Of Clusters}
  \label{fig:umass_topics_clus}
\end{subfigure}
\caption{We ran HDBScan for a range of minimum of clusters sizes as shown in Figure (a). The maximum mean topic coherance (UMASS) corresponds to a min cluster size of 200. In Figure (b), that is equivalent to 69 clusters (or topics). }
\label{fig:test}
\end{figure*}

\pagebreak\subsection{A2. Results Appendix}

\begin{table}[!h]
\centering
\begin{tabular}{l|l}
\hline
\rowcolor[HTML]{9B9B9B} 
\textbf{Topic ID} & \textbf{Generated Topic Name}       \\ \hline
13                & codeine\_lean\_syrup\_promethazine  \\ \hline
\rowcolor[HTML]{EFEFEF} 
40                & methadone\_quantity\_dose\_on       \\ \hline
61                & sober\_life\_addiction\_you         \\ \hline
\rowcolor[HTML]{EFEFEF} 
67                & foil\_smoke\_tar\_smoking           \\ \hline
14                & gabapentin\_clonidine\_gaba\_for    \\ \hline
\rowcolor[HTML]{EFEFEF} 
19                & constipation\_stomach\_pee\_this    \\ \hline
2                 & sex\_testosterone\_sex drive\_drive \\ \hline
\rowcolor[HTML]{EFEFEF} 
42                & oxy\_withdrawals\_day\_quantity     \\ \hline
33                & pain\_doctor\_my\_pain management   \\ \hline
\end{tabular}
\caption{Examples of generated topic names based on top keywords in cluster documents. Topic 13, 14 and 40 relate to questions about specific drugs like codeine, methadone and gabapentin. On the other hand, topics 2 and 19 talk about side effects. Some topics like topic 33 and 42 talk about withdrawal symptoms and pain management respectively. }
\label{table:topicinterpret}
\end{table}

\begin{table*}
\centering
\resizebox{1.8\columnwidth}{!}{%
\begin{tabular}{r|
>{\columncolor[HTML]{B7E1CD}}l |
>{\columncolor[HTML]{A6E3B6}}l |l}
\hline
\multicolumn{1}{l|}{\cellcolor[HTML]{C0C0C0}\textbf{Topic Id}} &
  \cellcolor[HTML]{C0C0C0}\textbf{Topic} &
  \cellcolor[HTML]{C0C0C0}\textbf{Group} &
  \cellcolor[HTML]{C0C0C0}\textbf{Percentage} \\ \hline
1  & { dark web, onion, vendors}                        & drug sales                             & \cellcolor[HTML]{FFCE93}23.25\% \\
2  & { issues with dealer}                              & drug sales                             & 0.59\%                          \\
3  & { finding a plug}                                  & drug sales                             & 0.22\%                          \\
4  & { suboxone dosing}                                 & addiction treatment                    & \cellcolor[HTML]{FFCE93}5.19\%  \\
5  & { methadone dosage}                                & addiction treatment                    & 1.19\%                          \\
6  & { community check ins}                             & addiction treatment                    & 0.69\%                          \\
7  & { soberity}                                        & addiction treatment                    & 0.93\%                          \\
8  & { take homes}                                      & addiction treatment                    & 0.25\%                          \\
9  & { narcan}                                          & addiction treatment                    & 0.24\%                          \\
10 & { detox}                                           & addiction treatment                    & 0.22\%                          \\
11 & { methadone + suboxone}                            & addiction treatment                    & 0.2\%                           \\
12 & { naloxone and naltrexone questions}               & addiction treatment                    & 0.2\%                           \\
13 & { sublocade shot}                                  & addiction treatment                    & 0.19\%                          \\
14 & { methadone clinics}                               & addiction treatment                    & 0.16\%                          \\
15 & { suboxone generics}                               & addiction treatment                    & 0.16\%                          \\
16 & { quitting}                                        & addiction treatment                    & 0.15\%                          \\
17 & { suboxone providers and insurance}                & addiction treatment                    & 0.13\%                          \\
18 & { codeine and promethazine}                        & specific drug                          & 5.14\%                          \\
19 & { tramadol questions}                              & specific drug                          & 1.25\%                          \\
20 & { gabapentin questions}                            & specific drug                          & 0.7\%                           \\
21 & { safe dosage of morphine}                         & specific drug                          & 0.6\%                           \\
22 & { tianeptine}                                      & specific drug                          & 0.52\%                          \\
23 & { safe dosage of oxy}                              & specific drug                          & 0.44\%                          \\
24 & { how much hydrocodone}                            & specific drug                          & 0.37\%                          \\
25 & { dilaudid impact and comparison to other opioids} & specific drug                          & 0.36\%                          \\
26 & { fentanyl and heroin}                                        & specific drug                          & 0.29\%                          \\
27 & { heroin}                                          & specific drug                          & 0.26\%                          \\
28 & { fentanyl patches}                                        & specific drug                          & 0.25\%                          \\
29 & { benzo and oxy}                                   & specific drug                          & 0.21\%                          \\
30 & { vivitrol treatment questions}                    & specific drug                          & 0.21\%                          \\
31 & { preferred opioid}                                & specific drug                          & 0.17\%                          \\
32 & { consuming poppy seeds}                           & specific drug                          & 0.17\%                          \\
33 & { urine drug test}                                 & drug testing                           & 1.52\%                          \\
34 & { kratom and drug tests}                           & drug testing                           & 0.33\%                          \\
35 & { injecting}                                       & drug use                               & 1.41\%                          \\
36 & { counterfeit vs real}                             & drug use                               & 1.25\%                          \\
37 & { foil smoking}                                    & drug use                               & 0.85\%                          \\
38 & { snorting}                                        & drug use                               & 0.69\%                          \\
39 & { filling medications}                             & drug use                               & 0.44\%                          \\
40 & { dosing and duration}                             & drug use                               & 0.32\%                          \\
41 & { dosage}                                          & drug use                               & 0.27\%                          \\
42 & { tolerance}                                       & drug use                               & 0.23\%                          \\
43 & { dosing}                                          & drug use                               & 0.19\%                          \\
44 & { overdosing}                                      & drug use                               & 0.15\%                          \\
45 & { tapering}                                        & withdrawal                             & 0.91\%                          \\
46 & { withdrawal}                                      & withdrawal                             & 0.59\%                          \\
47 & { withdrawals and quantity}                                      & withdrawal                             & 0.44\%                          \\
48 & { loperamide for withdrawal side effects}          & withdrawal                             & 0.25\%                          \\
49 & { Post-Acute Withdrawal Syndrome (PAWS)}           & withdrawal                             & 0.23\%                          \\
50 & { cold copping}                                    & withdrawal                             & 0.2\%                           \\
51 & { withdrawal symptoms timeline}                    & withdrawal                             & 0.2\%                           \\
52 & { cravings}                                        & withdrawal                             & 0.14\%                          \\
53 & { anyone in {[}location{]}}                        & lifestyle                              & 0.81\%                          \\
54 & { loved ones}                                      & lifestyle                              & 0.71\%                          \\
55 & { what to listen to and watch while high}          & lifestyle                              & 0.29\%                          \\
56 & { flying on planes with prescription and nonprescription substances} & lifestyle      & 0.19\%                          \\
57 & { substance-induced constipation}                  & side effect                            & 0.6\%                           \\
58 & { sex drive and methadone}                         & side effect                            & 0.57\%                          \\
59 & { restless legs}                                   & side effect                            & 0.36\%                          \\
60 & { dreams and nightmares}                           & side effect                            & 0.24\%                          \\
61 & { hair loss from kratom}                           & side effect                            & 0.23\%                          \\
62 & { sweating}                                        & side effect                            & 0.21\%                          \\
63 & { substance-induced nausea and vomiting}           & side effect                            & 0.21\%                          \\
64 & { opioids and itching}                             & side effect                            & 0.21\%                          \\
65 & { vision or pupil effects}                         & side effect                            & 0.2\%                           \\
66 & { nodding out}                                     & side effect                            & 0.2\%                           \\
67 & { grapefruit and its effects on medications}       & side effect                            & 0.17\%                          \\
68 & { pain management}                                 & pain management                        & 0.41\%                          \\
69 & { question about {[}x{]}}                          & miscellaneous brief question fragments & 0.78\%                          \\ \hline
\end{tabular}
}
\caption{Topics categorized into groups by public health experts based on top 100 relevant posts in every topic. The top 2 topics are marked in orange.
}
\label{tab:topicgroups}
\end{table*}

%% file: main_icwsm.bbl
\begin{thebibliography}{111}
\providecommand{\natexlab}[1]{#1}

\bibitem[{Asaduzzaman et~al.(2013)Asaduzzaman, Mashiyat, Roy, and Schneider}]{asaduzzaman}
Asaduzzaman, M.; Mashiyat, A.~S.; Roy, C.~K.; and Schneider, K.~A. 2013.
\newblock Answering Questions about Unanswered Questions of Stack Overflow.
\newblock In \emph{Proceedings of the 10th Working Conference on Mining Software Repositories}, MSR '13, 97–100. IEEE Press.
\newblock ISBN 9781467329361.

\bibitem[{Baker, Wade, and Walsh(2020)}]{baker2020covid19}
Baker, S.~A.; Wade, M.; and Walsh, M.~J. 2020.
\newblock <? covid19?> the challenges of responding to misinformation during a pandemic: content moderation and the limitations of the concept of harm.
\newblock \emph{Media International Australia}, 177(1): 103--107.

\bibitem[{Balsamo et~al.(2023)Balsamo, Bajardi, De~Francisci~Morales, Monti, and Schifanella}]{Balsamo_Bajardi_De_Francisci_Morales_Monti_Schifanella_2023}
Balsamo, D.; Bajardi, P.; De~Francisci~Morales, G.; Monti, C.; and Schifanella, R. 2023.
\newblock The Pursuit of Peer Support for Opioid Use Recovery on Reddit.
\newblock \emph{Proc. ICWSM}.

\bibitem[{Balsamo et~al.(2021)Balsamo, Bajardi, Salomone, and Schifanella}]{routePatterns}
Balsamo, D.; Bajardi, P.; Salomone, A.; and Schifanella, R. 2021.
\newblock Patterns of Routes of Administration and Drug Tampering for Nonmedical Opioid Consumption: Data Mining and Content Analysis of Reddit Discussions.
\newblock \emph{J Med Internet Res}, 23(1): e21212.

\bibitem[{Barua, Thomas, and Hassan(2014)}]{barua}
Barua, A.; Thomas, S.; and Hassan, A.~E. 2014.
\newblock What are developers talking about? An analysis of topics and trends in Stack Overflow.
\newblock \emph{Empirical Software Engineering}, 19.

\bibitem[{Baumgartner et~al.(2020)Baumgartner, Zannettou, Keegan, Squire, and Blackburn}]{baumgartner2020pushshift}
Baumgartner, J.; Zannettou, S.; Keegan, B.; Squire, M.; and Blackburn, J. 2020.
\newblock The Pushshift Reddit Dataset.
\newblock arXiv:2001.08435.

\bibitem[{Bell and Strang(2020)}]{bell2020medication}
Bell, J.; and Strang, J. 2020.
\newblock Medication treatment of opioid use disorder.
\newblock \emph{Biological psychiatry}, 87(1): 82--88.

\bibitem[{Beyer et~al.(2020)Beyer, Macho, Di~Penta, and Pinzger}]{beyer2020}
Beyer, S.; Macho, C.; Di~Penta, M.; and Pinzger, M. 2020.
\newblock What kind of questions do developers ask on Stack Overflow? A comparison of automated approaches to classify posts into question categories.
\newblock \emph{Empirical Software Engineering}, 25(3): 2258--2301.

\bibitem[{Bird, Klein, and Loper(2009)}]{nltk}
Bird, S.; Klein, E.; and Loper, E. 2009.
\newblock \emph{Natural language processing with Python: analyzing text with the natural language toolkit}.
\newblock " O'Reilly Media, Inc.".

\bibitem[{Bonnevie et~al.(2021)Bonnevie, Lloyd, Rosenberg, Williams, Goldbarg, and Smyser}]{bonnevie2021layla}
Bonnevie, E.; Lloyd, T.~D.; Rosenberg, S.~D.; Williams, K.; Goldbarg, J.; and Smyser, J. 2021.
\newblock Layla’s Got You: Developing a tailored contraception chatbot for Black and Hispanic young women.
\newblock \emph{Health Education Journal}, 80(4): 413--424.

\bibitem[{Bonnevie et~al.(2020)Bonnevie, Rosenberg, Kummeth, Goldbarg, Wartella, and Smyser}]{bonnevie2020using}
Bonnevie, E.; Rosenberg, S.~D.; Kummeth, C.; Goldbarg, J.; Wartella, E.; and Smyser, J. 2020.
\newblock Using social media influencers to increase knowledge and positive attitudes toward the flu vaccine.
\newblock \emph{Plos one}, 15(10): e0240828.

\bibitem[{Bunting et~al.(2021{\natexlab{a}})Bunting, Frank, Arshonsky, Bragg, Friedman, and Krawczyk}]{bunting2021socially}
Bunting, A.~M.; Frank, D.; Arshonsky, J.; Bragg, M.~A.; Friedman, S.~R.; and Krawczyk, N. 2021{\natexlab{a}}.
\newblock Socially-supportive norms and mutual aid of people who use opioids: An analysis of Reddit during the initial COVID-19 pandemic.
\newblock \emph{Drug and alcohol dependence}.

\bibitem[{Bunting et~al.(2021{\natexlab{b}})Bunting, Frank, Arshonsky, Bragg, Friedman, and Krawczyk}]{opioidcovid}
Bunting, A.~M.; Frank, D.; Arshonsky, J.; Bragg, M.~A.; Friedman, S.~R.; and Krawczyk, N. 2021{\natexlab{b}}.
\newblock {{S}ocially-supportive norms and mutual aid of people who use opioids: {A}n analysis of {R}eddit during the initial {C}{O}{V}{I}{D}-19 pandemic}.
\newblock \emph{Drug Alcohol Depend}, 222: 108672.

\bibitem[{Cavazos-Rehg et~al.(2018)Cavazos-Rehg, Zewdie, Krauss, and Sowles}]{browniehigh}
Cavazos-Rehg, P.~A.; Zewdie, K.; Krauss, M.~J.; and Sowles, S.~J. 2018.
\newblock {"{N}o {H}igh {L}ike a {B}rownie {H}igh": {A} {C}ontent {A}nalysis of {E}dible {M}arijuana {T}weets}.
\newblock \emph{Am J Health Promot}, 32(4): 880--886.

\bibitem[{{Centers for Disease Control and Prevention, National Center for Injury Prevention and Control}(2021)}]{cdcopioid:2021}
{Centers for Disease Control and Prevention, National Center for Injury Prevention and Control}. 2021.
\newblock {Understanding the Opioid Overdose Epidemic}.
\newblock \url{https://www.cdc.gov/overdose-prevention/about/understanding-the-opioid-overdose-epidemic.html}.
\newblock Accessed: 2022-09-01.

\bibitem[{Chancellor et~al.(2016)Chancellor, Lin, Goodman, Zerwas, and {De Choudhury}}]{chancellor2016quantifying}
Chancellor, S.; Lin, Z. J.~J.; Goodman, E.~L.; Zerwas, S.; and {De Choudhury}, M. 2016.
\newblock {Quantifying and Predicting Mental Illness Severity in Online Pro-Eating Disorder Communities}.
\newblock In \emph{CSCW}, 1169--1182.

\bibitem[{Chancellor et~al.(2019{\natexlab{a}})Chancellor, Nitzburg, Hu, Zampieri, and De~Choudhury}]{chancellor2019discovering}
Chancellor, S.; Nitzburg, G.; Hu, A.; Zampieri, F.; and De~Choudhury, M. 2019{\natexlab{a}}.
\newblock Discovering alternative treatments for opioid use recovery using social media.
\newblock In \emph{Proceedings of the 2019 CHI Conference on Human Factors in Computing Systems}, 1--15.

\bibitem[{Chancellor et~al.(2019{\natexlab{b}})Chancellor, Nitzburg, Hu, Zampieri, and De~Choudhury}]{alternateTreatments}
Chancellor, S.; Nitzburg, G.; Hu, A.; Zampieri, F.; and De~Choudhury, M. 2019{\natexlab{b}}.
\newblock Discovering Alternative Treatments for Opioid Use Recovery Using Social Media.
\newblock In \emph{Proceedings of the 2019 CHI Conference on Human Factors in Computing Systems}, CHI '19, 1–15. New York, NY, USA: Association for Computing Machinery.
\newblock ISBN 9781450359702.

\bibitem[{Chary et~al.(2017)Chary, Genes, Giraud-Carrier, Hanson, Nelson, and Manini}]{epidiopioids}
Chary, M.; Genes, N.; Giraud-Carrier, C.; Hanson, C.; Nelson, L.~S.; and Manini, A.~F. 2017.
\newblock {{E}pidemiology from {T}weets: {E}stimating {M}isuse of {P}rescription {O}pioids in the {U}{S}{A} from {S}ocial {M}edia}.
\newblock \emph{J Med Toxicol}, 13(4): 278--286.

\bibitem[{Ciccarone(2017)}]{ciccarone2017}
Ciccarone, D. 2017.
\newblock Fentanyl in the {US} heroin supply: A rapidly changing risk environment.
\newblock \emph{Int J Drug Policy}, 46: 107--111.

\bibitem[{Cummaudo et~al.(2020)Cummaudo, Vasa, Barnett, Grundy, and Abdelrazek}]{cummaudo}
Cummaudo, A.; Vasa, R.; Barnett, S.; Grundy, J.; and Abdelrazek, M. 2020.
\newblock Interpreting cloud computer vision pain-points: a mining study of stack overflow.
\newblock In Cleland-Huang, J.; and Marinov, D., eds., \emph{Proceedings - 2020 ACM/IEEE 42nd International Conference on Software Engineering, ICSE 2020}, 1584--1596. United States of America: Association for Computing Machinery (ACM).
\newblock International Conference on Software Engineering 2020, ICSE 2020 ; Conference date: 27-06-2020 Through 19-07-2020.

\bibitem[{Damiescu et~al.(2021)Damiescu, Banerjee, Lee, Paul, and Efferth}]{tramadolA}
Damiescu, R.; Banerjee, M.; Lee, D.~Y.; Paul, N.~W.; and Efferth, T. 2021.
\newblock Health(care) in the crisis: reflections in science and society on opioid addiction.
\newblock \emph{International Journal of Environmental Research and Public Health}, 18: 341.

\bibitem[{Das et~al.(2024)Das, Walker, Rajwal, Lakamana, Sumner, Mack, Kaczkowski, Sarker et~al.}]{das2024emerging}
Das, S.; Walker, D.; Rajwal, S.; Lakamana, S.; Sumner, S.~A.; Mack, K.~A.; Kaczkowski, W.; Sarker, A.; et~al. 2024.
\newblock Emerging trends of self-harm using sodium nitrite in an online suicide community: observational study using natural language processing analysis.
\newblock \emph{JMIR mental health}, 11(1): e53730.

\bibitem[{Ding et~al.(2024)Ding, Carik, Gunturi, Reyna, and Rho}]{Ding_2024}
Ding, X.; Carik, B.; Gunturi, U.~S.; Reyna, V.; and Rho, E. H.~R. 2024.
\newblock Leveraging Prompt-Based Large Language Models: Predicting Pandemic Health Decisions and Outcomes Through Social Media Language.
\newblock In \emph{Proceedings of the CHI Conference on Human Factors in Computing Systems}, CHI ’24, 1–20. ACM.

\bibitem[{Dirkson et~al.(2019)Dirkson, Verberne, Sarker, and Kraaij}]{lexnorm}
Dirkson, A.; Verberne, S.; Sarker, A.; and Kraaij, W. 2019.
\newblock Data-Driven Lexical Normalization for Medical Social Media.
\newblock \emph{Multimodal Technologies and Interaction}, 3(3).

\bibitem[{D’Agostino et~al.(2017)D’Agostino, Optican, Sowles, Krauss, Lee, and Cavazos-Rehg}]{d2017social}
D’Agostino, A.~R.; Optican, A.~R.; Sowles, S.~J.; Krauss, M.~J.; Lee, K.~E.; and Cavazos-Rehg, P.~A. 2017.
\newblock Social networking online to recover from opioid use disorder: A study of community interactions.
\newblock \emph{Drug and alcohol dependence}, 181: 5--10.

\bibitem[{ElSherief et~al.(2024)ElSherief, Sumner, Krishnasamy, Jones, Law, Kacha-Ochana, Schieber, and De~Choudhury}]{elsherief2024identification}
ElSherief, M.; Sumner, S.; Krishnasamy, V.; Jones, C.; Law, R.; Kacha-Ochana, A.; Schieber, L.; and De~Choudhury, M. 2024.
\newblock Identification of Myths and Misinformation About Treatment for Opioid Use Disorder on Social Media: Infodemiology Study.
\newblock \emph{JMIR Formative Research}, 8(1).

\bibitem[{ElSherief et~al.(2021)ElSherief, Sumner, Jones, Law, Kacha-Ochana, Shieber, Cordier, Holton, and De~Choudhury}]{misinfolargescale}
ElSherief, M.; Sumner, S.~A.; Jones, C.~M.; Law, R.~K.; Kacha-Ochana, A.; Shieber, L.; Cordier, L.; Holton, K.; and De~Choudhury, M. 2021.
\newblock Characterizing and Identifying the Prevalence of Web-Based Misinformation Relating to Medication for Opioid Use Disorder: Machine Learning Approach.
\newblock \emph{J Med Internet Res}, 23(12): e30753.

\bibitem[{Fleiss(1971)}]{fleiss1971measuring}
Fleiss, J.~L. 1971.
\newblock Measuring nominal scale agreement among many raters.
\newblock \emph{Psychological Bulletin}, 76(5): 378--382.

\bibitem[{Ford et~al.(2018)Ford, Lustig, Banks, and Parnin}]{ford2018}
Ford, D.; Lustig, K.; Banks, J.; and Parnin, C. 2018.
\newblock "We Don't Do That Here": How Collaborative Editing with Mentors Improves Engagement in Social Q\&A Communities.
\newblock 1--12.

\bibitem[{Fu and Oh(2018)}]{fu2018}
Fu, H.; and Oh, S. 2018.
\newblock Quality assessment of answers with user-identified criteria and data-driven features in social Q\&A.
\newblock \emph{Information Processing \& Management}, 56: 14--28.

\bibitem[{Fullwood, Kecojevic, and Basch(2016)}]{cannabinoids}
Fullwood, M.~D.; Kecojevic, A.; and Basch, C.~H. 2016.
\newblock {{E}xamination of {Y}ou{T}ube videos related to synthetic cannabinoids}.
\newblock \emph{Int J Adolesc Med Health}, 30(4).

\bibitem[{Gandhi et~al.(2022)Gandhi, Rouhani, Park, Urquhart, Allen, Morales, Green, and Sherman}]{gandhi2022alternative}
Gandhi, P.; Rouhani, S.; Park, J.~N.; Urquhart, G.~J.; Allen, S.~T.; Morales, K.~B.; Green, T.~C.; and Sherman, S.~G. 2022.
\newblock Alternative use of buprenorphine among people who use opioids in three US Cities.
\newblock \emph{Substance Abuse}, 43(1): 364--370.

\bibitem[{Glowacki, Glowacki, and Wilcox(2018)}]{opioidtextmining}
Glowacki, E.~M.; Glowacki, J.~B.; and Wilcox, G.~B. 2018.
\newblock A text-mining analysis of the public's reactions to the opioid crisis.
\newblock \emph{Substance Abuse}, 39(2): 129--133.
\newblock PMID: 28723265.

\bibitem[{Graves et~al.(2019)Graves, Sarker, Al-Garadi, Yang, Love, O’Connor, Gonzalez-Hernandez, and Perrone}]{buprenorphineTapering}
Graves, R.; Sarker, A.; Al-Garadi, M.~A.; Yang, Y.; Love, J.~S.; O’Connor, K.; Gonzalez-Hernandez, G.; and Perrone, J. 2019.
\newblock Effective buprenorphine use and tapering strategies: endorsements and insights by people in recovery from opioid use disorder on a reddit forum.

\bibitem[{Graves et~al.(2022)Graves, Perrone, Al-Garadi, Yang, Love, O'Connor, Gonzalez-Hernandez, and Sarker}]{buprenorphineNLP}
Graves, R.~L.; Perrone, J.; Al-Garadi, M.~A.; Yang, Y.-C.; Love, J.~S.; O'Connor, K.; Gonzalez-Hernandez, G.; and Sarker, A. 2022.
\newblock Thematic Analysis of Reddit Content About Buprenorphine-naloxone Using Manual Annotation and Natural Language Processing Techniques.
\newblock \emph{Journal of Addiction Medicine}.

\bibitem[{Grootendorst(2020)}]{bertopic}
Grootendorst, M. 2020.
\newblock BERTopic: Leveraging BERT and c-TF-IDF to create easily interpretable topics.

\bibitem[{Hanson et~al.(2013)Hanson, Cannon, Burton, and Giraud-Carrier}]{drugAbuseTwitter}
Hanson, C.~L.; Cannon, B.; Burton, S.; and Giraud-Carrier, C. 2013.
\newblock {{A}n exploration of social circles and prescription drug abuse through {T}witter}.
\newblock \emph{J Med Internet Res}, 15(9): e189.

\bibitem[{Harnett et~al.(2020)Harnett, Dines, Wood, Archer, and Dargan}]{coldWater2020}
Harnett, J.~T.; Dines, A.~M.; Wood, D.~M.; Archer, J. R.~H.; and Dargan, P.~I. 2020.
\newblock {{C}old water extraction of codeine/paracetamol combination products: a case series and literature review}.
\newblock \emph{Clin Toxicol (Phila)}, 58(2): 107--111.

\bibitem[{Hassan et~al.(2023)Hassan, Draman, Hassan, Ali, and Badrin}]{kratomMixing}
Hassan, H.; Draman, N.; Hassan, R.; Ali, N.; and Badrin, S. 2023.
\newblock The use of opioid in treating a patient with kratom use disorder: a case report.
\newblock \emph{Electronic Journal of General Medicine}, 20: em482.

\bibitem[{{HHS}(2021)}]{hhs}
{HHS}. 2021.
\newblock What is the U.S. Opioid Epidemic?
\newblock \url{https://www.hhs.gov/opioids/about-the-epidemic/index.html}.

\bibitem[{Hong et~al.(2021)Hong, Chen, Liu, Zhu, Yu, Ung, Chan, Hu, and Han}]{covidtrends}
Hong, S.; Chen, X.; Liu, X.; Zhu, H.; Yu, F.; Ung, C. O.~L.; Chan, W.~S.; Hu, H.; and Han, S. 2021.
\newblock National drug utilization trend of analgesics in china: an analysis of procurement data at 793 public hospitals from 2013 to 2018.
\newblock \emph{Journal of Pharmaceutical Policy and Practice}, 14.

\bibitem[{Hong et~al.(2020)Hong, Deng, Evans, and Wu}]{hong2020}
Hong, Z.; Deng, Z.; Evans, R.; and Wu, H. 2020.
\newblock Patient Questions and Physician Responses in a Chinese Health Q{\&}A Website: Content Analysis.
\newblock \emph{J Med Internet Res}, 22(4): e13071.

\bibitem[{Hu, Li, and Liang(2019)}]{hu-etal-2019-diachronic}
Hu, R.; Li, S.; and Liang, S. 2019.
\newblock Diachronic Sense Modeling with Deep Contextualized Word Embeddings: An Ecological View.
\newblock In Korhonen, A.; Traum, D.; and M{\`a}rquez, L., eds., \emph{Proceedings of the 57th Annual Meeting of the Association for Computational Linguistics}, 3899--3908. Florence, Italy: Association for Computational Linguistics.

\bibitem[{{Hugging Face}(2024)}]{huggingface_text_summ}
{Hugging Face}. 2024.
\newblock Hugging Face NLP Course: Chapter 7 - Text Summarization.
\newblock \url{https://huggingface.co/learn/nlp-course/en/chapter7/5}.
\newblock Accessed on 2024-05-15.

\bibitem[{Jenhani, Gouider, and Said(2019)}]{hybridsystem}
Jenhani, F.; Gouider, M.~S.; and Said, L.~B. 2019.
\newblock Hybrid System for Information Extraction from Social Media Text: Drug Abuse Case Study.
\newblock \emph{Procedia Computer Science}, 159: 688--697.
\newblock Knowledge-Based and Intelligent Information \& Engineering Systems: Proceedings of the 23rd International Conference KES2019.

\bibitem[{Katsuki, Mackey, and Cuomo(2015)}]{drugAbuseLinkTwit}
Katsuki, T.; Mackey, T.~K.; and Cuomo, R. 2015.
\newblock {{E}stablishing a {L}ink {B}etween {P}rescription {D}rug {A}buse and {I}llicit {O}nline {P}harmacies: {A}nalysis of {T}witter {D}ata}.
\newblock \emph{J Med Internet Res}, 17(12): e280.

\bibitem[{Kepner, Meacham, and Nobles(2022)}]{stigmaSources}
Kepner, W.; Meacham, M.~C.; and Nobles, A.~L. 2022.
\newblock Types and Sources of Stigma on Opioid Use Treatment and Recovery Communities on Reddit.
\newblock \emph{Substance Use \& Misuse}, 57(10): 1511--1522.
\newblock PMID: 35815614.

\bibitem[{Kim et~al.(2017)Kim, Marsch, Hancock, and Das}]{kim2017scaling}
Kim, S.~J.; Marsch, L.~A.; Hancock, J.~T.; and Das, A.~K. 2017.
\newblock Scaling up research on drug abuse and addiction through social media big data.
\newblock \emph{Journal of medical Internet research}, 19(10): e6426.

\bibitem[{Kim et~al.(2018)Kim, Okano, Osborne, Frank, Meana, and Castaneto}]{urineTest}
Kim, V.~J.; Okano, C.~K.; Osborne, C.~R.; Frank, D.~M.; Meana, C.~T.; and Castaneto, M.~S. 2018.
\newblock Can synthetic urine replace authentic urine to “beat” workplace drug testing?
\newblock \emph{Drug Testing and Analysis}, 11: 331--335.

\bibitem[{Krauss et~al.(2017)Krauss, Grucza, Bierut, and Cavazos-Rehg}]{tweetAnalysis}
Krauss, M.~J.; Grucza, R.~A.; Bierut, L.~J.; and Cavazos-Rehg, P.~A. 2017.
\newblock {"{G}et drunk. {S}moke weed. {H}ave fun.": {A} {C}ontent {A}nalysis of {T}weets {A}bout {M}arijuana and {A}lcohol}.
\newblock \emph{Am J Health Promot}, 31(3): 200--208.

\bibitem[{Langabeer et~al.(2020)Langabeer, Stotts, Cortez, Tortolero, and Champagne-Langabeer}]{langabeer2020geographic}
Langabeer, J.~R.; Stotts, A.~L.; Cortez, A.; Tortolero, G.; and Champagne-Langabeer, T. 2020.
\newblock Geographic proximity to buprenorphine treatment providers in the US.
\newblock \emph{Drug and alcohol dependence}, 213: 108131.

\bibitem[{Li et~al.(2021)Li, Du, Liao, Jiang, and Champagne-Langabeer}]{li2021demystifying}
Li, Z.; Du, X.; Liao, X.; Jiang, X.; and Champagne-Langabeer, T. 2021.
\newblock Demystifying the dark web opioid trade: content analysis on anonymous market listings and forum posts.
\newblock \emph{Journal of Medical Internet Research}, 23(2): e24486.

\bibitem[{Liang et~al.(2021)Liang, Chen, Bennett, and Yang}]{liang2021identifying}
Liang, O.~S.; Chen, Y.; Bennett, D.~S.; and Yang, C.~C. 2021.
\newblock Identifying self-management support needs for pregnant women with opioid misuse in online health communities: mixed methods analysis of web posts.
\newblock \emph{Journal of medical Internet research}, 23(2): e18296.

\bibitem[{Lin(2004)}]{rouge}
Lin, C.-Y. 2004.
\newblock {ROUGE}: A Package for Automatic Evaluation of Summaries.
\newblock In \emph{Text Summarization Branches Out}, 74--81. Barcelona, Spain: Association for Computational Linguistics.

\bibitem[{Linares-V\'{a}squez et~al.(2014)Linares-V\'{a}squez, Bavota, Di~Penta, Oliveto, and Poshyvanyk}]{linares}
Linares-V\'{a}squez, M.; Bavota, G.; Di~Penta, M.; Oliveto, R.; and Poshyvanyk, D. 2014.
\newblock How Do API Changes Trigger Stack Overflow Discussions? A Study on the Android SDK.
\newblock In \emph{Proceedings of the 22nd International Conference on Program Comprehension}, ICPC 2014, 83–94. New York, NY, USA: Association for Computing Machinery.
\newblock ISBN 9781450328791.

\bibitem[{Liu et~al.(2016)Liu, Tang, He, Zhou, and Yao}]{hierclust}
Liu, L.; Tang, L.; He, L.; Zhou, W.; and Yao, S. 2016.
\newblock An Overview of Hierarchical Topic Modeling.
\newblock In \emph{2016 8th International Conference on Intelligent Human-Machine Systems and Cybernetics (IHMSC)}, volume~01, 391--394.

\bibitem[{Love et~al.(2023)Love, Levine, Aldy, Brent, Krotulski, Logan, Vargas-Torres, Walton, Amaducci, Calello et~al.}]{love2023opioid}
Love, J.~S.; Levine, M.; Aldy, K.; Brent, J.; Krotulski, A.~J.; Logan, B.~K.; Vargas-Torres, C.; Walton, S.~E.; Amaducci, A.; Calello, D.; et~al. 2023.
\newblock Opioid overdoses involving xylazine in emergency department patients: a multicenter study.
\newblock \emph{Clinical Toxicology}, 61(3): 173--180.

\bibitem[{Lu et~al.(2019)Lu, Sridhar, Pandey, Hasan, and Mohler}]{lu2019investigate}
Lu, J.; Sridhar, S.; Pandey, R.; Hasan, M.~A.; and Mohler, G. 2019.
\newblock Investigate transitions into drug addiction through text mining of Reddit data.
\newblock In \emph{Proceedings of the 25th ACM SIGKDD International Conference on Knowledge Discovery \& Data Mining}, 2367--2375.

\bibitem[{Lynch et~al.(2015)Lynch, Shapiro, Coffa, Novak, and Kral}]{promethazine}
Lynch, K.~L.; Shapiro, B.~J.; Coffa, D.; Novak, S.~P.; and Kral, A.~H. 2015.
\newblock {{P}romethazine use among chronic pain patients}.
\newblock \emph{Drug Alcohol Depend}, 150: 92--97.

\bibitem[{MacLean et~al.(2015)MacLean, Gupta, Lembke, Manning, and Heer}]{maclean2015forum77}
MacLean, D.; Gupta, S.; Lembke, A.; Manning, C.; and Heer, J. 2015.
\newblock Forum77: An analysis of an online health forum dedicated to addiction recovery.
\newblock In \emph{Proceedings of the 18th ACM Conference on Computer Supported Cooperative Work \& Social Computing}, 1511--1526.

\bibitem[{Mamykina et~al.(2011)Mamykina, Manoim, Mittal, Hripcsak, and Hartmann}]{mamykina}
Mamykina, L.; Manoim, B.; Mittal, M.; Hripcsak, G.; and Hartmann, B. 2011.
\newblock Design Lessons from the Fastest Q\&amp;a Site in the West.
\newblock In \emph{Proceedings of the SIGCHI Conference on Human Factors in Computing Systems}, CHI '11, 2857–2866. New York, NY, USA: Association for Computing Machinery.
\newblock ISBN 9781450302289.

\bibitem[{Marraffa et~al.(2018)Marraffa, Stork, Hoffman, and Su}]{marraffa2018poison}
Marraffa, J.~M.; Stork, C.~M.; Hoffman, R.~S.; and Su, M.~K. 2018.
\newblock Poison control center experience with tianeptine: an unregulated pharmaceutical product with potential for abuse.
\newblock \emph{Clinical toxicology}, 56(11): 1155--1158.

\bibitem[{McInnes, Healy, and Astels(2017)}]{hdbscan}
McInnes, L.; Healy, J.; and Astels, S. 2017.
\newblock hdbscan: Hierarchical density based clustering.
\newblock \emph{Journal of Open Source Software}, 2(11): 205.

\bibitem[{McInnes, Healy, and Melville(2020)}]{umap}
McInnes, L.; Healy, J.; and Melville, J. 2020.
\newblock UMAP: Uniform Manifold Approximation and Projection for Dimension Reduction.
\newblock arXiv:1802.03426.

\bibitem[{McLaren et~al.(2023)McLaren, Jones, Noonan, Idaikkadar, and Sumner}]{mclaren2023trends}
McLaren, N.; Jones, C.~M.; Noonan, R.; Idaikkadar, N.; and Sumner, S.~A. 2023.
\newblock Trends in stigmatizing language about addiction: A longitudinal analysis of multiple public communication channels.
\newblock \emph{Drug and Alcohol Dependence}, 245: 109807.

\bibitem[{Nadi and Treude(2020)}]{nadi2020}
Nadi, S.; and Treude, C. 2020.
\newblock Essential Sentences for Navigating Stack Overflow Answers.
\newblock \emph{2020 IEEE 27th International Conference on Software Analysis, Evolution and Reengineering (SANER)}, 229--239.

\bibitem[{O’Donnell et~al.(2023)O’Donnell, Tanz, Miller, Dinwiddie, Wolff, Mital, Obiekwe, and Mattson}]{donnell2023}
O’Donnell, J.; Tanz, L.~J.; Miller, K.~D.; Dinwiddie, A.~T.; Wolff, J.; Mital, S.; Obiekwe, R.; and Mattson, C.~L. 2023.
\newblock Drug Overdose Deaths with Evidence of Counterfeit Pill Use — United States, July 2019–December 2021.
\newblock \emph{MMWR Morb Mortal Wkly Rep}.

\bibitem[{Papineni et~al.(2002)Papineni, Roukos, Ward, and Zhu}]{bleu}
Papineni, K.; Roukos, S.; Ward, T.; and Zhu, W.-J. 2002.
\newblock BLEU: A Method for Automatic Evaluation of Machine Translation.
\newblock In \emph{Proceedings of the 40th Annual Meeting on Association for Computational Linguistics}, ACL '02, 311–318. USA: Association for Computational Linguistics.

\bibitem[{Park, Conway, and Chen(2018)}]{reddittextmining}
Park, A.; Conway, M.; and Chen, A.~T. 2018.
\newblock Examining thematic similarity, difference, and membership in three online mental health communities from reddit: A text mining and visualization approach.
\newblock \emph{Computers in Human Behavior}, 78: 98--112.

\bibitem[{Pascali et~al.(2018)Pascali, Fais, Vaiano, Pigaiani, D'Errico, Furlanetto, Palumbo, and Bertol}]{drugTampering2018}
Pascali, J.~P.; Fais, P.; Vaiano, F.; Pigaiani, N.; D'Errico, S.; Furlanetto, S.; Palumbo, D.; and Bertol, E. 2018.
\newblock {{I}nternet pseudoscience: {T}esting opioid containing formulations with tampering potential}.
\newblock \emph{J Pharm Biomed Anal}, 153: 16--21.

\bibitem[{Peiper et~al.(2019)Peiper, Clarke, Vincent, Ciccarone, Kral, and Zibbell}]{peiper2019fentanyl}
Peiper, N.~C.; Clarke, S.~D.; Vincent, L.~B.; Ciccarone, D.; Kral, A.~H.; and Zibbell, J.~E. 2019.
\newblock Fentanyl test strips as an opioid overdose prevention strategy: findings from a syringe services program in the Southeastern United States.
\newblock \emph{International Journal of Drug Policy}, 63: 122--128.

\bibitem[{Peprah et~al.(2020)Peprah, Agyemang-Duah, Appiah-Brempong, Akwasi, and Morgan}]{tramadolCrisis}
Peprah, P.; Agyemang-Duah, W.; Appiah-Brempong, E.; Akwasi, A.~G.; and Morgan, A.~K. 2020.
\newblock ``With tramadol, I ride like a Jaguar'': a qualitative study of motivations for non-medical purpose tramadol use among commercial vehicle operators in Kumasi, Ghana.
\newblock \emph{Substance Abuse Treatment, Prevention, and Policy}, 15(1): 49.

\bibitem[{Peters et~al.(2016)Peters, Pontones, Hoover, Patel, Galang, Shields, Blosser, Spiller, Combs, Switzer et~al.}]{peters2016hiv}
Peters, P.~J.; Pontones, P.; Hoover, K.~W.; Patel, M.~R.; Galang, R.~R.; Shields, J.; Blosser, S.~J.; Spiller, M.~W.; Combs, B.; Switzer, W.~M.; et~al. 2016.
\newblock HIV infection linked to injection use of oxymorphone in Indiana, 2014--2015.
\newblock \emph{New England Journal of Medicine}, 375(3): 229--239.

\bibitem[{Pieper et~al.(2009)Pieper, Templin, Kirsner, and Birk}]{Pieper2009}
Pieper, B.; Templin, T.~N.; Kirsner, R.~S.; and Birk, T.~J. 2009.
\newblock Impact of injection drug use on distribution and severity of chronic venous disorders.
\newblock \emph{Wound Repair Regen}, 17(4): 485--491.

\bibitem[{Procaci et~al.(2019)Procaci, Siqueira, Pereira~Nunes, and Nurmikko-Fuller}]{procaci2019}
Procaci, T.~B.; Siqueira, S. W.~M.; Pereira~Nunes, B.; and Nurmikko-Fuller, T. 2019.
\newblock Experts and Likely to Be Closed Discussions in Question and Answer Communities: An Analytical Overview.
\newblock \emph{Comput. Hum. Behav.}, 92(C): 519–535.

\bibitem[{Razaque et~al.(2021)Razaque, Valiyev, Alotaibi, Alotaibi, Amanzholova, and Alotaibi}]{coviddarkweb}
Razaque, A.; Valiyev, B.; Alotaibi, B.; Alotaibi, M.; Amanzholova, S.; and Alotaibi, A. 2021.
\newblock Influence of covid-19 epidemic on dark web contents.
\newblock \emph{Electronics}, 10: 2744.

\bibitem[{Rehurek and Sojka(2011)}]{gensim}
Rehurek, R.; and Sojka, P. 2011.
\newblock Gensim--python framework for vector space modelling.
\newblock \emph{NLP Centre, Faculty of Informatics, Masaryk University, Brno, Czech Republic}, 3(2).

\bibitem[{Reimers and Gurevych(2019)}]{reimers-2019-sentence-bert}
Reimers, N.; and Gurevych, I. 2019.
\newblock Sentence-BERT: Sentence Embeddings using Siamese BERT-Networks.
\newblock In \emph{Proceedings of the 2019 Conference on Empirical Methods in Natural Language Processing}. Association for Computational Linguistics.

\bibitem[{Romano et~al.(2024)Romano, Sharif, Basak, Gatto, and Preum}]{Romano_Sharif_Basak_Gatto_Preum_2024}
Romano, W.; Sharif, O.; Basak, M.; Gatto, J.; and Preum, S.~M. 2024.
\newblock Theme-Driven Keyphrase Extraction to Analyze Social Media Discourse.
\newblock \emph{Proceedings of the International AAAI Conference on Web and Social Media}, 18(1): 1315--1327.

\bibitem[{Rosen and Shihab(2016)}]{rosen}
Rosen, C.; and Shihab, E. 2016.
\newblock What Are Mobile Developers Asking about? A Large Scale Study Using Stack Overflow.
\newblock \emph{Empirical Softw. Engg.}, 21(3): 1192–1223.

\bibitem[{Rubya and Yarosh(2017)}]{rubya2017video}
Rubya, S.; and Yarosh, S. 2017.
\newblock Video-mediated peer support in an online community for recovery from substance use disorders.
\newblock In \emph{Proceedings of the 2017 ACM Conference on Computer Supported Cooperative Work and Social Computing}, 1454--1469.

\bibitem[{Rudasill et~al.(2019)Rudasill, Sanaiha, Mardock, Khoury, Xing, Antonios, McKinnell, and Benharash}]{rudasill2019clinical}
Rudasill, S.~E.; Sanaiha, Y.; Mardock, A.~L.; Khoury, H.; Xing, H.; Antonios, J.~W.; McKinnell, J.~A.; and Benharash, P. 2019.
\newblock Clinical outcomes of infective endocarditis in injection drug users.
\newblock \emph{Journal of the American College of Cardiology}, 73(5): 559--570.

\bibitem[{Sadowski, Stolee, and Elbaum(2015)}]{sadowksi}
Sadowski, C.; Stolee, K.~T.; and Elbaum, S. 2015.
\newblock How Developers Search for Code: A Case Study.
\newblock In \emph{Proceedings of the 2015 10th Joint Meeting on Foundations of Software Engineering}, ESEC/FSE 2015, 191–201. New York, NY, USA: Association for Computing Machinery.
\newblock ISBN 9781450336758.

\bibitem[{SAMHSA(2021)}]{samhsa2020}
SAMHSA. 2021.
\newblock 2020 {N}ational {S}urvey of {D}rug {U}se and {H}ealth ({NSDUH}) {R}eleases.

\bibitem[{Sammut and Webb(2010)}]{tfidf}
Sammut, C.; and Webb, G.~I., eds. 2010.
\newblock \emph{TF--IDF}, 986--987.
\newblock Boston, MA: Springer US.
\newblock ISBN 978-0-387-30164-8.

\bibitem[{Santos et~al.(2020)Santos, Burghardt, Lerman, and Helic}]{santos2020}
Santos, T.; Burghardt, K.; Lerman, K.; and Helic, D. 2020.
\newblock Can Badges Foster a More Welcoming Culture on Q\&amp;A Boards?
\newblock \emph{Proceedings of the International AAAI Conference on Web and Social Media}, 14(1): 969--973.

\bibitem[{Sarker, DeRoos, and Perrone(2020)}]{sarker2020mining}
Sarker, A.; DeRoos, A.; and Perrone, J. 2020.
\newblock Mining social media for prescription medication abuse monitoring: a review and proposal for a data-centric framework.
\newblock \emph{Journal of the American Medical Informatics Association}, 27(2): 315--329.

\bibitem[{Sarker et~al.(2015)Sarker, Ginn, Nikfarjam, O’Connor, Smith, Jayaraman, Upadhaya, and Gonzalez}]{sarker2015utilizing}
Sarker, A.; Ginn, R.; Nikfarjam, A.; O’Connor, K.; Smith, K.; Jayaraman, S.; Upadhaya, T.; and Gonzalez, G. 2015.
\newblock Utilizing social media data for pharmacovigilance: a review.
\newblock \emph{Journal of biomedical informatics}, 54: 202--212.

\bibitem[{Sarker et~al.(2016)Sarker, O’connor, Ginn, Scotch, Smith, Malone, and Gonzalez}]{sarker2016social}
Sarker, A.; O’connor, K.; Ginn, R.; Scotch, M.; Smith, K.; Malone, D.; and Gonzalez, G. 2016.
\newblock Social media mining for toxicovigilance: automatic monitoring of prescription medication abuse from Twitter.
\newblock \emph{Drug safety}, 39(3): 231--240.

\bibitem[{Sequeira et~al.(2019)Sequeira, Gayen, Ganguly, Dandapat, and Chandra}]{sequeira2019large}
Sequeira, R.; Gayen, A.; Ganguly, N.; Dandapat, S.~K.; and Chandra, J. 2019.
\newblock A large-scale study of the Twitter follower network to characterize the spread of prescription drug abuse tweets.
\newblock \emph{IEEE Transactions on Computational Social Systems}, 6(6): 1232--1244.

\bibitem[{Sharif et~al.(2023)Sharif, Basak, Parvin, Scharfstein, Bradham, Borodovsky, Lord, and Preum}]{sharif2023characterizing}
Sharif, O.; Basak, M.; Parvin, T.; Scharfstein, A.; Bradham, A.; Borodovsky, J.~T.; Lord, S.~E.; and Preum, S.~M. 2023.
\newblock Characterizing Information Seeking Events in Health-Related Social Discourse.
\newblock arXiv:2308.09156.

\bibitem[{Shen et~al.(2020)Shen, Jia, Shen, and Dou}]{Shen2020HelpingTI}
Shen, X.; Jia, A.~L.; Shen, S.; and Dou, Y. 2020.
\newblock Helping the Ineloquent Farmers: Finding Experts for Questions With Limited Text in Agricultural Q\&A Communities.
\newblock \emph{IEEE Access}, 8: 62238--62247.

\bibitem[{Shleifer and Rush(2020)}]{bertsum}
Shleifer, S.; and Rush, A.~M. 2020.
\newblock Pre-trained Summarization Distillation.
\newblock \emph{CoRR}, abs/2010.13002.

\bibitem[{Spadaro et~al.(2022)Spadaro, Sarker, Hogg-Bremer, Love, O'Donnell, Nelson, and Perrone}]{spadaro2022reddit}
Spadaro, A.; Sarker, A.; Hogg-Bremer, W.; Love, J.~S.; O'Donnell, N.; Nelson, L.~S.; and Perrone, J. 2022.
\newblock Reddit discussions about buprenorphine associated precipitated withdrawal in the era of fentanyl.
\newblock \emph{Clinical Toxicology}, 60(6): 694--701.

\bibitem[{Spencer, Miniño, and Warner(2022)}]{ahmad2021provisional}
Spencer, M.~R.; Miniño, A.~M.; and Warner, M. 2022.
\newblock Drug Overdose Deaths in the United States, 2001-2021.
\newblock \emph{National Center for Health Statistics}, 12.

\bibitem[{Stieglitz et~al.(2018)Stieglitz, Mirbabaie, Ross, and Neuberger}]{socialmediachallenges}
Stieglitz, S.; Mirbabaie, M.; Ross, B.; and Neuberger, C. 2018.
\newblock Social media analytics – Challenges in topic discovery, data collection, and data preparation.
\newblock \emph{International Journal of Information Management}, 39: 156--168.

\bibitem[{Sue and Fiellin(2021)}]{sue2021bringing}
Sue, K.~L.; and Fiellin, D.~A. 2021.
\newblock Bringing harm reduction into health policy-combating the overdose crisis.
\newblock \emph{N Engl J Med}, 384(19): 1781--1783.

\bibitem[{Turc et~al.(2019)Turc, Chang, Lee, and Toutanova}]{bert-mini}
Turc, I.; Chang, M.; Lee, K.; and Toutanova, K. 2019.
\newblock Well-Read Students Learn Better: The Impact of Student Initialization on Knowledge Distillation.
\newblock \emph{CoRR}, abs/1908.08962.

\bibitem[{Vindrola-Padros and Johnson(2020)}]{vindrola2020rapid}
Vindrola-Padros, C.; and Johnson, G.~A. 2020.
\newblock Rapid techniques in qualitative research: a critical review of the literature.
\newblock \emph{Qualitative health research}, 30(10): 1596--1604.

\bibitem[{Volkow and McLellan(2016)}]{volkow2016opioid}
Volkow, N.~D.; and McLellan, A.~T. 2016.
\newblock Opioid abuse in chronic pain—misconceptions and mitigation strategies.
\newblock \emph{New England Journal of Medicine}, 374(13): 1253--1263.

\bibitem[{Wadden et~al.(2021)Wadden, August, Li, and Althoff}]{wadden2021effect}
Wadden, D.; August, T.; Li, Q.; and Althoff, T. 2021.
\newblock The effect of moderation on online mental health conversations.
\newblock \emph{ICWSM}.

\bibitem[{Wang, Chen, and Hassan(2018)}]{fastAns}
Wang, S.; Chen, T.-H.; and Hassan, A.~E. 2018.
\newblock Understanding the Factors for Fast Answers in Technical Q\&amp;A Websites.
\newblock \emph{Empirical Softw. Engg.}, 23(3): 1552–1593.

\bibitem[{Wang(2021)}]{wang2021}
Wang, Y. 2021.
\newblock {The price of being polite: politeness, social status, and their joint impacts on community Q\&A efficiency}.
\newblock \emph{Journal of Computational Social Science}, 4(1): 101--122.

\bibitem[{Weld, Zhang, and Althoff(2022)}]{weld2022makes}
Weld, G.; Zhang, A.~X.; and Althoff, T. 2022.
\newblock What Makes Online Communities ‘Better’? Measuring Values, Consensus, and Conflict across Thousands of Subreddits.
\newblock In \emph{Proceedings of the International AAAI Conference on Web and Social Media}, volume~16, 1121--1132.

\bibitem[{Williams et~al.(2022)Williams, Mauro, Feng, Wilson, Cruz, Olfson, Crystal, Samples, and Chiodo}]{williams2022non}
Williams, A.~R.; Mauro, C.~M.; Feng, T.; Wilson, A.; Cruz, A.; Olfson, M.; Crystal, S.; Samples, H.; and Chiodo, L. 2022.
\newblock Non--prescribed buprenorphine preceding treatment intake and clinical outcomes for opioid use disorder.
\newblock \emph{Journal of Substance Abuse Treatment}, 108770.

\bibitem[{Wolf et~al.(2020)Wolf, Debut, Sanh, Chaumond, Delangue, Moi, Cistac, Rault, Louf, Funtowicz, Davison, Shleifer, von Platen, Ma, Jernite, Plu, Xu, Scao, Gugger, Drame, Lhoest, and Rush}]{huggingface}
Wolf, T.; Debut, L.; Sanh, V.; Chaumond, J.; Delangue, C.; Moi, A.; Cistac, P.; Rault, T.; Louf, R.; Funtowicz, M.; Davison, J.; Shleifer, S.; von Platen, P.; Ma, C.; Jernite, Y.; Plu, J.; Xu, C.; Scao, T.~L.; Gugger, S.; Drame, M.; Lhoest, Q.; and Rush, A.~M. 2020.
\newblock HuggingFace's Transformers: State-of-the-art Natural Language Processing.
\newblock arXiv:1910.03771.

\bibitem[{Yao et~al.(2020)Yao, Rashidian, Dong, Duanmu, Rosenthal, and Wang}]{yao2020detection}
Yao, H.; Rashidian, S.; Dong, X.; Duanmu, H.; Rosenthal, R.~N.; and Wang, F. 2020.
\newblock Detection of suicidality among opioid users on reddit: machine learning--based approach.
\newblock \emph{Journal of medical internet research}, 22(11): e15293.

\bibitem[{Zeng et~al.(2019)Zeng, Dubreuil, LaRochelle, Lu, Wei, Choi, Lei, and Zhang}]{tramadolB}
Zeng, C.; Dubreuil, M.; LaRochelle, M.~R.; Lu, N.; Wei, J.; Choi, H.~K.; Lei, G.; and Zhang, Y. 2019.
\newblock Association of Tramadol With {All-Cause} Mortality Among Patients With Osteoarthritis.
\newblock \emph{JAMA}, 321(10): 969--982.

\bibitem[{Zhang et~al.(2019{\natexlab{a}})Zhang, Wang, Chen, Zou, and Hassan}]{zhang2019stackoverflow}
Zhang, H.; Wang, S.; Chen, T.; Zou, Y.; and Hassan, A.~E. 2019{\natexlab{a}}.
\newblock An Empirical Study of Obsolete Answers on Stack Overflow.
\newblock \emph{CoRR}, abs/1903.12282.

\bibitem[{Zhang et~al.(2019{\natexlab{b}})Zhang, Lu, Phang, and Zhang}]{Zhang2019ScientificKC}
Zhang, Y.; Lu, T.; Phang, C.~W.; and Zhang, C. 2019{\natexlab{b}}.
\newblock Scientific Knowledge Communication in Online Q\&A Communities: Linguistic Devices as a Tool to Increase the Popularity and Perceived Professionalism of Knowledge Contribution.
\newblock \emph{J. Assoc. Inf. Syst.}, 20: 3.

\end{thebibliography}
